\documentclass{article}
\usepackage[utf8]{inputenc}

\usepackage{pgf,tikz}
\usepackage{pgfplots}       
\usetikzlibrary{intersections}
\usepgfplotslibrary{colormaps,external, fillbetween}
\usepgfplotslibrary{groupplots}
\usepgfplotslibrary{statistics}
\usetikzlibrary{fit,calc}

\pgfplotsset{compat=1.18}

\usepackage[pdfencoding=auto]{hyperref}

\usepackage{bm}

\usepackage{geometry}
\usepackage{graphicx}

\usepackage[citestyle=authoryear, backend=bibtex, sorting=nyt, natbib=true, url=false, doi=false]{biblatex}
\usepackage{amsfonts}
\usepackage{amsmath}
\usepackage{amssymb}

\usepackage{subcaption}

\usepackage{rotating}

\usepackage[utf8]{inputenc}
\usepackage[T1]{fontenc}
\usepackage{mathpazo} 

\usepackage{xcolor}
\usepackage[skins]{tcolorbox}

\tcbset{commonstyle/.style={boxrule=0pt,sharp corners,enhanced jigsaw,nobeforeafter,boxsep=0pt,left=\fboxsep,right=\fboxsep}}

\newtcolorbox{mycolorbox}[1][]{commonstyle,#1}

\definecolor{colorblue}{RGB}{97,189,252}
\definecolor{colorred}{RGB}{252,189,97}

\usepackage{booktabs}

\usepackage[parfill]{parskip} 

\usepackage{authblk}

\usepackage{algorithm}
\usepackage{algpseudocode}

\providecommand{\keywords}[1]
{
  \small	
  \textbf{\textit{Keywords---}} #1
}

\newcommand{\Mod}[1]{\ \mathrm{mod}\ #1}

\algnewcommand{\IfThen}[2]{
	\State \algorithmicif\ #1\ \algorithmicthen\ #2}
\algnewcommand{\LineComment}[1]{\State \(\triangleright\) #1}

\algnewcommand{\InlineFor}[2]{
	\State \algorithmicfor\ #1\ \algorithmicendfor\ #2}

 \makeatletter
\AfterEndEnvironment{algorithm}{\let\@algcomment\relax}
\AtEndEnvironment{algorithm}{\kern2pt\hrule\relax\vskip3pt\@algcomment}
\let\@algcomment\relax
\newcommand\algcomment[1]{\def\@algcomment{\footnotesize#1}}

\renewcommand\fs@ruled{\def\@fs@cfont{\bfseries}\let\@fs@capt\floatc@ruled
  \def\@fs@pre{\hrule height.8pt depth0pt \kern2pt}%
  \def\@fs@post{}%
  \def\@fs@mid{\kern2pt\hrule\kern2pt}%
  \let\@fs@iftopcapt\iftrue}
\makeatother

\title{Routing One Million Customers in a Handful of Minutes}

\author[1]{Luca Accorsi}
\author[2, 3]{Daniele Vigo}

\date{\nolinkurl{accorsi@google.com}\\
      \nolinkurl{daniele.vigo@unibo.it}}

\affil[1]{Google}
\affil[2]{Department of Electrical, Electronic and Information Engineering ``G. Marconi'', University of Bologna, Italy}
\affil[3]{CIRI ICT, University of Bologna, Italy}

\begin{document}

\maketitle

\begin{abstract}
In this paper, we propose a new dataset of Capacitated Vehicle Routing Problem instances which are up to two orders of magnitude larger than those in the currently used benchmarks. Despite these sizes might not have an immediate application to real-world logistic scenarios, we believe they could foster fresh new research efforts on the design of effective and efficient algorithmic components for routing problems. We provide computational results for such instances by running FILO2, an adaptation of the FILO algorithm proposed in \citet{filo}, designed to handle extremely large-scale CVRP instances. Solutions for such instances are obtained by using an a standard personal computer in a considerably short computing time, thus showing the effectiveness of the acceleration and pruning techniques already proposed in FILO. Finally, results of FILO2 on well-known literature instances show that the newly introduced changes improve the overall scalability of the approach with respect to the previous FILO design.
\end{abstract}


\section{Introduction} \label{sec:intro}
The Capacitated Vehicle Routing Problem (CVRP, see \cite{VRPBOOK}) is a challenging and practically relevant combinatorial optimization problem that, since its proposal in \citet{dantzig59}, has attracted a huge research effort for the development of efficient solution methods. Algorithmic advancements and increase in raw computing power made possible the heuristic resolution of larger and larger problem instances. 
Most of the existing state-of-the-art algorithms for the CVRP are evaluated by using the well-known $\mathbb{X}$ dataset proposed by \citet{UCHOA2017845}, which include instances with up to 1,000 customers, as has also happened for the recent \citet{DIMACS} implementation challenge. To the best of our knowledge, the $\mathbb{B}$ dataset proposed by \citet{arnold2019xxl} contains the largest literature instances, including up to  thirty thousand customers. 
Such large instances 
may have practical applications, indeed, the authors based their construction on real-world parcel distribution problems in Belgium. 


Devising algorithms able to scale up to the size of the $\mathbb{B}$ instances and far beyond, other than being a mere research exercise, may promote new research ideas also applicable to instances with a smaller scales. For this reason, in this paper we propose what we call extremely large (XXL) instances, with up to one million customers, as a new additional  benchmark for future research on large-scale CVRP. We also provide benchmark results obtained by an evolution of the FILO method proposed in \citet{filo} explicitly designed to handle XXL instances.

The CVRP is defined on an undirected complete graph $G = (V, E)$ where $V$ is the vertex set and $E$ is the edge set. The set of vertices $V$ is partitioned into $V = \{0\} \cup V_c$ where $0$ is the depot and $V_c = \{1, 2, \ldots, N\}$ is a set of $N$ customers. A cost $c_{ij}$ is associated with each edge $(i, j) \in E$ and we assume that the cost matrix $\bm{c}$ satisfies the triangle inequality. For a subset $V' \subseteq V$, we identify with $\mathcal{N}_i^k(V')$ the ordered list of the $k$ nearest neighbor vertices $j \in V'$ of vertex $i$ with respect to the cost matrix $\bm{c}$. Whenever $k = |V'|$, we refer to the ordered list of neighbors of $i$ by omitting the apex $k$, i.e., $\mathcal{N}_i(V')$. Each customer $i \in V_c$ requires an integer quantity $q_i > 0$ of homogeneous goods from the depot, and for the depot we have $q_0 = 0$.
An unlimited fleet of homogeneous vehicles with capacity $Q$ is located at the depot, available to serve the customers. Recalling that a Hamiltonian circuit is a closed cycle visiting a set of customers exactly once, a CVRP solution $S$ is made up of a number $|S|$ of Hamiltonian circuits, called routes, starting from the depot, visiting a subset of customers, and coming back to the depot. A solution is feasible if all customers are visited exactly once and the load of all the used vehicles does not exceed the capacity $Q$. The cost of a solution $S$ is given by the sum of the costs of the edges defining the routes of $S$ and the goal of the CVRP is to find the feasible solution with the minimum cost.

Successful approaches to large-scale CVRP instances typically make use of acceleration and pruning techniques to speed up their optimization process. More specifically, these techniques often support the execution of the local search, which is typically one of the most computationally expensive procedure of a heuristic solution approach.
Among the most popular and successful acceleration and pruning techniques we have Granular Neighborhoods (GNs), Static Move Descriptors (SMDs) and Sequential Search (SS). 
GNs, originally proposed in \citet{DBLP:journals/informs/TothV03}, make use of the fact that for most local search operators, neighboring solutions can be generated by a local search move uniquely determined by an arc $(i, j) \in E$, called the \textit{move generator}. Thus, instead of exploring a complete neighborhood, a sparsification rule is used to specify which arcs are used to generate local search moves. The computational complexity of such a neighborhood exploration can be arbitrarily reduced according to how move generators are defined. The right amount of move generators, defined by proper sparsification rules, experimentally shown to provide a good tradeoff between local search execution time and solution quality.
In the SMD framework proposed by \citet{Zachariadis:2010:SRC:1805350.1805623}, the classic for-loop exploration of neighborhoods is replaced with a structured inspection of data structures, called SMDs, that identify a move and its effect, in terms of cost change to the current solution. The use of SMDs makes neighborhood exploration much more efficient, because whenever a change is applied to the current solution it is possible to evaluate a smaller neighborhood affected by the newly applied change, by updating a subset of the existing SMDs, rather than performing a complete reevaluation of the whole neighborhood.
GNs and SMDs have been successfully employed in several algorithms, such as \citet{DBLP:journals/informs/TothV03, Zachariadis:2010:SRC:1805350.1805623, DBLP:journals/eor/SchneiderSV17, xsttrp, filo}. Finally, the SS proposed by \citet{IRNICH20062405} decomposes a local search move into basic blocks called \textit{partial moves}. The execution of these partial moves can be aborted if certain conditions are met, thus pruning in advance a non-promising local search move. This technique was successfully used in \citet{arnold2019kgls, arnold2019xxl}.

Decomposition techniques are another tool for dealing with large-scale instances. Their idea relates to the divide-and-conquer principle. First, the original problem is split into several subproblems which are generally considerably smaller than then original problem. The subproblems are then solved efficiently, and their solutions are combined together to produce an overall solution to the original problem. 
\citet{santini21} provide a systematic characterization of decomposition techniques for vehicle routing heuristics and show how the Hybrid Genetic Search, a state-of-the-art algorithm for a broad class of vehicle routing problems (see \citet{vidal12} and \citet{vidal22}), could benefit from them when approaching large-scale instances.

Algorithm FILO proposed in \citet{filo} takes, on the other hand, a different approach to cope with large-scale instances. It employs the above mentioned GNs and SMDs, but instead of explicitly decomposing the instance, it relies on the localized optimization of a very limited and dynamically defined solution area. This is accomplished through a heuristic pruning technique called Selective Vertex Caching (SVC) that consists in using a cache-like data structure to keep track of a limited set of vertices involved in recent solution changes. 

More specifically, we can analyze the SVC functioning in FILO by taking a look at the neighbor solution generation occurring during its core optimization phase. First, a neighbor solution $S'$ is obtained by copying the current working solution $S$. At this point the SVC of $S'$ is empty since no action has ever been performed before directly to $S'$. A shaking procedure executed in a ruin-and-recreate fashion is applied to $S'$. During the shaking, a set of vertices is removed and then reinserted with a greedy algorithm. Every removal and insertion causes affected vertices to be inserted into the SVC. A subsequent local search procedure is applied by exploring local search moves generated by move generators $(i, j)$ such that at least one between $i$ and $j$ belong to the SVC. 
The net effect is that the neighbor solution generation is largely independent from the actual instance size but rather more affected by several algorithm parameters such as the SVC maximum capacity, the ruin intensity and the number of move generators considered from every vertex used during the local search. Indeed, results proposed in \citet{filo} show that approximately the same computing time to solve instances with $100$ customers is required to solve instances with ten times more customers.  

We shall however also note that a major drawback of an SVC approach compared to an explicit decomposition into independent subproblems is given by the inherent sequentiality of the former. Indeed, if several independent subproblems resulting from the decomposition can be easily solved in parallel without synchronization efforts, the implicit decomposition offered by the SVC makes parallelization a much more challenging task.

The contributions of this paper are the following:
\begin{itemize}
    \item We introduce a dataset of XXL instances obtained by projecting to the Euclidean plane the geographical coordinates of existing addresses in various Italian regions. The dataset contains instances with a number of customers ranging from $20$ thousand to $1$ million.
    \item We develop FILO2 as an evolution of FILO to show the scalability properties of the techniques already proposed in \citet{filo} as well as some limitations of the original algorithm. In fact, as discussed in Section \ref{sec:solution_approach}, as we move towards much larger instances, given the localization of the neighbor solution generation, other procedures become the actual bottleneck of the approach and some changes are required for the overall algorithm to retain its efficiency.
    \item We provide computational results of FILO2 on existing literature instances, showing that it matches the quality of FILO and greatly improves its scalability, and on the newly introduced instances, showing how FILO2 is able to tackle such XXL instances within an ordinary computing system. 
\end{itemize}

The paper is structured as follows. Section \ref{sec:solution_approach} describes FILO2, highlighting the differences with respect to FILO. Section \ref{sec:computaional_testing} provides the experimental results, and in Section \ref{sec:analysis} we provide some analysis on the behaviour of the algorithm. Finally, the Appendix contains supplemental material about detailed computational results and statistical analysis.

\section{Solution Approach} \label{sec:solution_approach}
The proposed solution approach, called FILO2, is an evolution of FILO designed to handle XXL instances of the CVRP. 
Our goal is indeed showing that the original FILO algorithm, that was among the best performing algorithms on very large-scale instances during the 12th DIMACS implementation challenge (see \citet{DIMACS}), is already capable of tackling much larger instances provided that relatively minor adjustments are performed.

In the following sections, we first provide an introductory description of FILO, then we move to a detailed description of FILO2, by highlighting the main changes with respect to the original algorithm along with their motivation.

\subsection{The FILO Metaheuristic} \label{sec:filo}
Algorithm FILO is a randomized, effective, and efficient algorithm specifically designed for large-scale instances of the CVRP. Algorithm \ref{algorithm:filo} and the following paragraphs provide an high-level overview of the approach. Furthermore, Table \ref{tab:filo-params} provides a reference for most of the parameters used in FILO and mentioned in this overview. A thorough description of the original approach is beyond the scope of this paper, we thus refer the interested reader to \citet{filo} for more details about its design. 
\begin{algorithm}
	\footnotesize
	\caption{High-level FILO structure}\label{algorithm:filo}
	\algcomment{Input parameters: instance $\mathcal{I}$, seed $s$}
	\begin{algorithmic}[1]
		\Procedure{filo}{$\mathcal{I}, s$}
		\State $\mathcal{R} \gets \Call{RandomEngine}{s}$
		\State $S \gets \Call{Construction}{ }$
		\State $k \gets \Call{GreedyRoutesEstimate}{\mathcal{I}}$
		\IfThen{$|S| > k$}{$S \gets \Call{RouteMin}{S, \mathcal{R}}$}
		\State $S \gets \Call{CoreOpt}{S, \mathcal{R}}$
        \State \Return $S$
		\EndProcedure
	\end{algorithmic}
\end{algorithm}

The algorithm begins with a construction phase that builds an initial feasible solution with a restricted version of the savings algorithm (see \citet{cw64}) proposed by \citet{arnold2019xxl}. More precisely, instead of computing all possible savings, which grow quadratically with the instance size, only a linear number of them are computed. Such a limited savings algorithm has experimentally shown to generate, for medium to large-scale instances, initial solutions with a quality comparable to that of a complete savings algorithm. Once an initial solution is available, an improvement phase is executed. During this phase a route minimization procedure is optionally performed if it is estimated that more routes than necessary are employed. The estimation of the ideal number of routes $k$ is performed by greedily solving the bin packing problem associated with the CVRP instance with a simple first-fit algorithm (see, e.g., Chapter 8 of \citet{KP}). Then, a core optimization procedure, which is the main procedure of FILO, is executed. 

Both the route minimization and the core optimization procedures are based on the Iterated Local Search paradigm (ILS, \citet{Lourenco2003}), in which shaking and local search applications interleave for a fixed number of iterations. In both procedures, the shaking is performed in a \textit{ruin-and-recreate} fashion (see \citet{Schrimpf}), and the local search makes use of GNs to heuristically prune unpromising moves, and SMDs to avoid unnecessary neighborhood reevaluations. Moreover, the SVC is always used to keep the optimization mainly localized to the specific area that has just been affected by the shaking procedure.
In particular, a vertex $i$ is inserted into the SVC whenever an operation involving $i$ is performed on the current solution. As an example, the removal of vertex $i$ from its route $r_i$ would insert into the SVC $i$ itself, as well as the predecessor $\pi_i$ and the successor $\sigma_i$ of $i$ in $r_i$. The maximum cache capacity is defined by a parameter $C$ and is managed with a first-in-first-out policy: whenever a new vertex should be inserted when the cache is full, the least recently inserted vertex is removed.

\begin{table}[b!]
	\footnotesize
	\centering
	\caption{Parameters of Algorithm FILO mentioned in the overview.\label{tab:filo-params}}
	\begin{tabular}{p{3cm} lp{11.5cm}}
	\toprule
	\\
	\multicolumn{3}{l}{Initial solution definition} \\
	\midrule
	$n_{cw} = 100$        && Number of neighbors considered in the savings computation of the construction phase.\\
	$\Delta_{RM} = 10^3$ && Maximum number of route minimization iterations.\\
	\\
	\multicolumn{3}{l}{Granular neighborhood}\\
	\midrule
	$n_{gs} = 25$        && Number of neighbors considered by the sparsification rule.\\
	$\gamma_{base} = 0.25$ && Base sparsification factor, i.e., at least $n_{gs} \cdot \gamma_{base}$ move generators are always considered for every vertex.\\
	$\delta = 0.5$ && Reduction factor used in the definition of the fraction of non-improving iterations performed before increasing a sparsification parameter.\\
	$\lambda = 2$ && Sparsification increment factor.\\
	\\
	\multicolumn{3}{l}{Core optimization} \\
	\midrule
	$\Delta_{CO} = 10^5, 10^6$ && Number of core optimization iterations for short and long runs, respectively.\\
	$\mathcal{T}_0$, $\mathcal{T}_f$ && Initial and final simulated annealing temperature.\\
	$C = 50$             && Maximum capacity of the SVC.\\
	$\omega_{base} = \lceil \ln{|V|} \rceil$ && Initial shaking intensity.\\
	\\
	\bottomrule
	\end{tabular}
	\end{table}

Several local search operators such as CROSS-exchange (see \citet{taillard1997tabu}), 2-opt (see \citet{Reinelt1994TheTS}) variants, and ejection chain (see \citet{GLOVER1996223}) are explored in a variable neighborhood descent fashion (VND, see \citet{MLADENOVIC19971097}) and concur in making a powerful optimization action. 

The route minimization procedure may visit infeasible solutions to perform its route compacting action, while the core optimization procedure explores only the feasible space and achieves its diversification by means of an effective simulated annealing acceptance rule (SA, see \citet{Kirkpatrick671}).

\subsubsection{Route Minimization}
\begin{algorithm}[!ht]
	\footnotesize
	\caption{Route minimization procedure}
	\algcomment{Input parameters: instance $\mathcal{I}$, solution $S^*$, route estimate $k$, random engine $\mathcal{R}$}
	\label{algorithm:routemin}
	\begin{algorithmic}[1]
		\Procedure{RouteMin}{$\mathcal{I}, S^*, k, \mathcal{R}$}
        \State $S \gets S^*, \mathcal{P} \gets 1, L \gets [\;]$ \label{algorithm:routemin:1}
		\For{$n \gets 1 \text{ to } \Delta_{RM}$}
		    \State $(r_i, r_j) \gets \Call{PickPairOfRoutes}{S, \mathcal{R}}$ \label{algorithm:routemin:2}
		    \State $S \gets S \smallsetminus r_i \smallsetminus r_j$ \label{algorithm:routemin:3}
			\State $L \gets [L, \Call{CustomersOf}{r_i, r_j}]$
		    \State $\bar{L} = [\;]$
		    \For{$i \in \Call{Ordered}{L, \mathcal{I}, \mathcal{R}}$} \label{algorithm:routemin:4}
		        \State $p \gets \Call{BestInsertionPositionInExistingRoutes}{i,S}$
		        \If{$p \not= none$}
                    \State $S \gets \Call{Insert}{i, p, S}$
                \Else
                    \If{$|S| < k \lor U(0, 1) > \mathcal{P}$} \label{algorithm:routemin:7}
                        \State $S \gets \Call{BuildSingleCustomerRoute}{i, S}$
                    \Else
                        \State $\bar{L} \gets [\bar{L}, i]$
                    \EndIf \label{algorithm:routemin:8}
		        \EndIf
		    \EndFor \label{algorithm:routemin:5}
		    \State $L \gets \bar{L}$
		    \State $S \gets \Call{LocalSearch}{S, \mathcal{R}}$ \label{algorithm:routemin:6}

		    \If{$|L| = 0 \land (\Call{Cost}{S} < \Call{Cost}{S^*} \lor (\Call{Cost}{S} = \Call{Cost}{S^*} \land |S| < |S^*|))$} \label{algorithm:routemin:9}
                \State $S^* \gets S$
                \IfThen{$|S^*| \leq k$}{ \Return $S^{*}$}
		    \EndIf \label{algorithm:routemin:10}
		    \State $\mathcal{P} \gets \Call{Decrease}{\mathcal{P}}$
            \IfThen{$\Call{Cost}{S} > \Call{Cost}{S^*}$}{$S \gets S^*$}
		\EndFor
		\State \Return $S^{*}$
		\EndProcedure
	\end{algorithmic}
\end{algorithm}
The route minimization procedure, whose pseudocode is shown in Algorithm \ref{algorithm:routemin}, is optionally applied to the initial solution $S$ generated by the construction phase. 

Initially, the working solution $S$ is set to be equal to the current best found solution $S^*$ originally obtained by the construction phase (line \ref{algorithm:routemin:1}). 
Then, for a prefixed number $\Delta_{RM}$ of iterations, the following main steps are performed: (i) a pair of routes $(r_i, r_j)$ and their associated customers are removed from $S$ (lines \ref{algorithm:routemin:2} and \ref{algorithm:routemin:3}), (ii) customers are greedily reinserted, after been ordered according to certain criteria, in the resulting partial solution $S$ in the position that minimizes their insertion cost (lines \ref{algorithm:routemin:4} to \ref{algorithm:routemin:5}), and finally, (iii) a local search is applied to $S$ (line \ref{algorithm:routemin:6}). 

The key procedure idea is that whenever an insertion position for a customer cannot be found within the existing routes, i.e., all the possible insertion positions would violate the route feasibility by requiring a total route load greater than the maximum vehicle capacity $Q$, such customer can probabilistically remain unserved for the current iteration (lines \ref{algorithm:routemin:7} to \ref{algorithm:routemin:8}). When this happens, in the subsequent iteration step (ii) would also consider these unserved customers.
The probability of leaving a customer unserved decreases over the iterations by following an exponential schedule, so that in later iterations customers are less likely to be left unserved. 
Whenever, after the local search execution in step (iii), the solution $S$ results to be better than $S^*$, the best found solution $S^*$ is replaced with $S$ (lines \ref{algorithm:routemin:9} to \ref{algorithm:routemin:10}). 
The comparison between $S$ and $S^*$ happens through the evaluation of their costs and number of routes. In particular, $S$ is considered to be better than $S^*$ whenever it has a lower cost, or in the unlikely case in which it has the same cost but a lower number of routes. This criterion follows the CVRP definition where a better objective function is always preferred over a lower number of routes. Nevertheless, the route minimization procedure experimentally proved to be both fast, requiring only few seconds for medium and large instances, and on average beneficial in terms of final solution quality for those instances to which it was applied. 

\subsubsection{Core Optimization}
\begin{algorithm}
	\footnotesize
	\caption{Core optimization procedure}
	\label{algorithm:coreopt}
	\algcomment{Input parameters: solution $S^*$, random engine $\mathcal{R}$}
	\begin{algorithmic}[1]
		\Procedure{CoreOpt}{$S^*, \mathcal{R}$}
        \State $S \gets S^*, \mathcal{T} \gets \mathcal{T}_0$ \label{algorithm:coreopt:1}
		\State $\bm{\omega} \gets (\omega_1, \omega_2, \ldots, \omega_{|V_c|}), \omega_i \gets \omega_{base} \; \forall i \in V_c$ 
		\State $\bm{\gamma} \gets (\gamma_0, \gamma_1, \ldots, \gamma_{|V_c|}), \gamma_i \gets \gamma_{base} \; \forall i \in V$ 
		\For{$n \gets 1 \text{ to } \Delta_{CO}$}
		    \State $S' \gets S$ \label{algorithm:coreopt:2}
			\State $S' \gets \Call{Shaking}{S', \mathcal{R}, \bm{\omega}}$
            \State $S' \gets \Call{LocalSearch}{S', \mathcal{R}}$ \label{algorithm:coreopt:3}
            \If{$\Call{Cost}{S'} < \Call{Cost}{S^*}$}
                \State $S^* \gets S'$ \label{algorithm:coreopt:4}
                \State $\Call{ResetSparsificationFactors}{\bm{\gamma}}$ \label{algorithm:coreopt:6}

            \Else
                \State $\Call{UpdateSparsificationFactors}{\bm{\gamma}}$ \label{algorithm:coreopt:7}
            \EndIf
            \State $\Call{UpdateShakingParameters}{\bm{\omega}, S', S, \mathcal{R}}$ 
            \IfThen{$\Call{AcceptNeighbor}{S, S', \mathcal{T}}$}{$S \gets S'$} \label{algorithm:coreopt:5}
            \State $\mathcal{T} \gets c \cdot \mathcal{T}$
		\EndFor

		\State \Return $S^{*}$
		\EndProcedure
	\end{algorithmic}
\end{algorithm}
The core optimization, whose pseudocode is shown in Algorithm \ref{algorithm:coreopt}, is an iterative procedure performed for a number $\Delta_{CO}$ iterations that is the main actor of the solution improvement in FILO.

As an initial setup step, the working solution $S$ is set to be equal to current best solution $S^*$ defined by the initial construction phase and possibly optimized by the route minimization procedure (line \ref{algorithm:coreopt:1}). 
Then, during every iteration, a neighbor solution $S'$ is generated through the application of shaking and local search procedures to solution $S$ (lines \ref{algorithm:coreopt:2} to \ref{algorithm:coreopt:3}). Solution $S^*$ is replaced by $S'$ if the latter has a lower cost (line \ref{algorithm:coreopt:4}). Moreover, $S'$ may replace $S$ based on a classical simulated annealing criterion (line \ref{algorithm:coreopt:5}).

The shaking procedure is performed in a ruin-and-recreate fashion. More precisely, the ruin step removes a number  $\omega_i$ of customers, starting from a randomly selected seed customer $i$. This is achieved by performing a random walk rooted in $i$, of length $\omega_i$ on the partial graph representing the solution and by removing every visited customer. 
The recreate step reinserts the removed customers in the position that minimizes their insertion cost after they have been sorted according to certain criteria (i.e., either by increasing or decreasing distance from the depot, or by decreasing demand value).
The vector $\bm{\omega} = (\omega_1, \ldots, \omega_{|V_c|})$ defines a per-customer ruin intensity, whose value is initially set to a $\omega_{base} = \lceil \ln{|V|} \rceil$ and then iteratively updated, for all customers $i$ involved in the ruin step, according to the quality of the resulting neighbor solution $S'$ with respect to the source solution $S$.

The SVC is empty at beginning of every iteration before the ruin application (line \ref{algorithm:coreopt:2}) since no action has been performed to $S'$ yet. The shaking application causes some vertices involved in the ruin-and-recreate procedures to be inserted into the SVC. The actual cached vertices are defined by the order in which vertices are processed and by the SVC maximum capacity. 
Because the explored local search moves are identified by move generators $(i, j)$ such that at least a vertex between $i$ or $j$ is included in the SVC, the local search will mainly optimize the solution area affected by the shaking. This makes every core optimization iteration extremely efficient and largely independent from the actual instance size.

The local search intensity is guided by a vertex-wise sparsification level $\bm{\gamma} = (\gamma_0, \ldots, \gamma_{|V_c|})$ that is a percentage determining, for each vertex $i \in V$, the number of move generators used during a local search application involving $i$.
The complete set of move generators is defined to be $T = \cup_{i \in V} \{(i, j), (j, i) \in E: j \in \mathcal{N}^{n_{gs}}_i(V \smallsetminus \{i\})\}$, that is, for every vertex $i$, the arcs connecting $i$ to the $n_{gs}$ neighbors $j$. Parameters $\gamma_i$ are used to select a fraction, in increasing arc cost, of these move generators for every vertex $i$.
At the beginning of the procedure, these parameters are initialized to $\gamma_i = \gamma_{base}, i \in V$. 
Whenever a solution $S'$ is found to improve the best known solution $S^*$, parameters $\gamma_i$ for vertices $i$ involved the last local search application are reset to $\gamma_{base}$ (line \ref{algorithm:coreopt:6}). 
On the other hand, if a vertex $i$ is involved in a certain number of consecutive unsuccessful local search applications, the value of $\gamma_i$ is set to be $\gamma_i = min\{\gamma_i \cdot \lambda, 1\}$, where $\lambda$ is an increment factor (line \ref{algorithm:coreopt:7}).
Vertices $i$ for which $\gamma_i$ is updated are those into the SVC after the local search application.

\subsection{The FILO2 Metaheuristic} \label{sec:filo2}
Algorithm FILO2 extends FILO to support XXL instances. In particular, it mostly removes the nonlinear (and some linear) procedures still existing within the original algorithm, and limits the memory occupation of its data structures. The following sections describe in detail the implemented changes, by  highlighting the differences between FILO and FILO2.

\subsubsection{Cost Matrix Storage} \label{sec:cost-matrix}
Before starting the actual optimization process, the original FILO algorithm performs an initial instance preprocessing during which some data structures are computed once, to be then used throughout the whole algorithm. In particular, during this initial instance preprocessing, all costs $c_{ij}$ for $i, j \in V$ are computed and stored in the cost matrix $\bm{c}$. Such a quadratic memory occupation is no longer feasible when moving to XXL instances. For this reason in FILO2, the cost matrix $\bm{c}$ is never explicitly stored but costs are computed on-demand whenever required. 

In addition, every solution object keeps track of the costs of the arcs composing it, i.e., if $j$ is served after $i$, the cost $c_{ij}$ is associated with vertex $j$ and identified by $\hat{c}_j$. This allows to access in constant-time most of the costs used during the algorithm procedures. As an example, consider a relocate local search move removing vertex $i$ from its current position and inserting it before vertex $j$. 
By denoting with $\pi_i$ and $\sigma_i$ the predecessor and successor of vertex $i$ respectively, the solution cost variation of this change is given by $c_{\pi_i \sigma_i} - c_{\pi_i i} - c_{i \sigma_i} + c_{\pi_j i} + c_{i j} - c_{\pi_j j}$. Half of the required costs, and more precisely those describing pairs of consecutive vertices $c_{\pi_i i}$, $c_{i \sigma_i}$, and $c_{\pi_j j}$ can be retrieved in constant-time from the current solution by accessing $\hat{c}_i$, $\hat{c}_{\sigma_i}$, and $\hat{c}_j$. The decision of associating the cost $c_{ij}$ to vertex $j$ is completely arbitrary, indeed $c_{ij}$ could have been associated with $i$ as well. Moreover, associating to each vertex $i$ the cost $c_{\pi_ii}$ and not also the cost $c_{i\sigma_i}$ is enough for the access pattern of FILO2. 

Furthermore, since every local search operator is designed according to the GN paradigm, a local search move identified by move generator $(i, j)$ will always try to insert arc $(i, j)$ in solution. The arc costs $c_{ij}$ can thus be associated with move generator $(i, j)$ allowing for an additional constant-time cost retrieval. Thus, in the above relocate example, only two out of six costs need to be computed in practice. 
Finally, since the CVRP is associated with a symmetric cost matrix, a single copy between $c_{ij}$ and $c_{ji}$ can be associated with both move generators $(i, j)$ and $(j, i)$ halving the memory requirements.

\subsubsection{Neighbor Lists Storage} \label{sec:neighbor_lists}
In FILO, during the above-mentioned initial preprocessing, also neighbor lists $\mathcal{N}_i(V)$ for every $i \in V$ are fully populated, i.e., for every vertex $i \in V$, vertices $j \in V$ are sorted according to their distance to $i$ and stored in increasing order. As for the cost matrix storage, storing all neighbors for all vertices would require a quadratic memory occupation and it is thus not feasible from a memory perspective when solving XXL instances. Moreover, completely sorting all vertices to populate these lists is also not acceptable from a computing time perspective since it would require $|V|^2 log |V|$ operations.

We shall, however, note that such neighbor lists are very important data structures that are used in key procedures of the algorithm. In particular, they support the following ones.
\begin{itemize}
    \item \textbf{The construction phase based on the savings algorithm.} As mentioned in Section \ref{sec:filo}, not all savings are computed, but only a limited number of them. More precisely, for every customer $i \in V_c$, only savings $s_{ij}$ with $j \in \mathcal{N}_i^{n_{cw}}(V_c)$ are computed. Therefore, for every customer $i$ at least the $n_{cw}$ nearest customers $j$ need to be known to maintain the same construction algorithm proposed in FILO. The parameter $n_{cw}$ was originally set to $100$ because the experimental analysis showed that such value achieves a good trade-off between computing time and initial solution quality. 
    \item \textbf{The ruin step of the core optimization and route minimization shaking procedures.} During the ruin step of the core optimization procedure, a random walk of prefixed length is performed by starting from a random seed customer. Whenever a customer is visited during the walk, it is removed from the solution to be later reinserted during the recreate step (see Section \ref{sec:recreate}). A random walk ending at customer $i$ of route $r_i$ may be extended either by moving within the same route, thus visiting the predecessor or the successor of $i$, or by jumping to a neighbor route possibly different from $r_i$. Such a neighbor route is identified by scanning customers $j \in \mathcal{N}_{i}(V_c)$ and searching for a customer $j$ served by route $r_j$ possibly such that $r_i \not= r_j$.
    
    Similarly, during the ruin step of the route minimization procedure, two close routes are selected, destroyed, and all their customers removed from the current solution. The first route is identified by randomly selecting a seed customer $i$, then the second route is again identified by searching customers $j \in \mathcal{N}_i(V_c)$ for one such that $r_i \not= r_j$.
    \item \textbf{The move generators definition for GNs.} The set of move generators $T$ used to perform the local search procedures is defined to contain all arcs connecting a vertex $i \in V$ to its $n_{gs} = 25$ nearest vertices, i.e., $T = \cup_{i \in V} \{(i, j), (j, i) \in E: j \in \mathcal{N}^{n_{gs}}_i(V \smallsetminus \{i\})\}$.
\end{itemize}
In FILO2, the neighbor lists are restricted to contain a limited number $n_{nn}$ of vertices, i.e., for every vertex $i$, only $\mathcal{N}_i^{n_{nn}}(V)$ is computed and stored during the initial instance preprocessing. These vertices are efficiently identified by exploiting a $kd$-tree data structure built on top of vertex coordinates (see \citet{bentley1990k}). As described in Section \ref{sec:parameter_tuning} and analyzed in Section \ref{sec:analysis_instance_preprocessing}, the value for $n_{nn}$ is set to be $n_{gs} < n_{cw} \leq n_{nn} \ll |V|$ to keep both the construction phase and the move generators definition as in the original algorithm. As far as the ruin step of the route minimization and core optimization procedures is concerned, the value of $n_{nn}$ is large enough to make sure these procedures are almost always able to find nearest vertices satisfying the required conditions, e.g., belonging to a route different from the current one. In the unlikely case in which this is not possible, the ruin procedure is simply prematurely concluded, e.g., the random walk is not extended in the core optimization procedure or a single route is removed in the route minimization procedure.

\subsubsection{Recreate Strategy} \label{sec:recreate}
In the original FILO algorithm, the recreate step of the core optimization and route minimization shaking procedures takes all the customers removed by the ruin step, sorts them according to a certain criterion (distance from the depot or customer demand), and greedily reinserts them in the best possible position, i.e., the position that currently minimizes the insertion cost. When such a position cannot be found in the existing routes because all candidate positions would violate the capacity constraints, a new single-customer route is created in the core optimization procedure or the customer may probabilistically remain unserved in the route minimization procedure.

When moving to XXL instances, fully scanning a partial solution searching for the best insertion position is  computationally too expensive. Thus, in FILO2, for every removed customer $i$, only routes $r_j$ that serve neighbor customers $j \in \mathcal{N}_i^{n_{nn}}(V)$ are scanned searching for candidate insertion positions.

\subsubsection{Arc Cost Sampling} \label{sec:arc-cost-sampling}
The search trajectory of FILO is ruled by a neighbor acceptance criterion based on the SA paradigm. 
In particular, during the core optimization procedure shaking and local search are applied to the current working solution $S$ to generate a neighbor solution $S'$. Solution $S'$ will replace solution $S$ in the subsequent core optimization iteration if $\Call{cost}{S'} < \Call{cost}{S} + \mathcal{T} \cdot \ln{U(0, 1)}$, where $\mathcal{T}$ is the current temperature. The value of $\mathcal{T}$ is bounded between an initial and final temperature $\mathcal{T}_0$ and $\mathcal{T}_f$, respectively, and it is updated at every core optimization iteration according to $\mathcal{T} = c \cdot \mathcal{T}$, where $c = (\mathcal{T}_f / \mathcal{T}_0)^{(1 / \Delta_{CO})}$ and $\Delta_{CO}$ is the total number of core optimization phase iterations. The initial and final temperatures $\mathcal{T}_0$ and $\mathcal{T}_f$ are defined to be proportional to the average cost $d$ of an arc of the instance. More precisely $\mathcal{T}_0 = 0.1 \cdot d$ with $d = \sum_{i, j \in V : i < j} c_{ij} / (|V| \cdot (|V| - 1) / 2)$, and $\mathcal{T}_f = 0.01 \cdot \mathcal{T}_0$. This was done to adapt the algorithm to instances having considerably different arc costs. 

In FILO2, we replace $d$ with an estimation $\tilde{d}$ based on a limited random sampling of a number of $|V|$ arcs. We observed a minor difference between the exact and estimated average arc cost for most of the instances. Moreover, this difference becomes even smaller for the SA initial temperature since the average arc cost is down-scaled by a factor of ten.

\subsubsection{Solution Copies} \label{sec:do-undo-lists}
Every core optimization iteration of FILO starts from an initial solution $S$ and produces a neighbor solution $S'$ obtained by the shaking and local search applications. 
As described in Section \ref{sec:arc-cost-sampling}, solution $S'$ could replace the current working solution $S$ if the SA acceptance criterion is satisfied (line \ref{algorithm:coreopt:4} of Algorithm \ref{algorithm:coreopt}). Moreover, $S'$ could replace the best solution $S^*$ in case it improves it (line \ref{algorithm:coreopt:5} of Algorithm \ref{algorithm:coreopt}).
Because shaking and local search applications are extremely efficient, the simple copy of solutions for XXL instances, despite being a linear operation in the instance size, is an expensive procedure that when performed hundreds of thousands of times during the algorithm execution, becomes a critical bottleneck.
In addition, given the locality of the optimization of FILO, the differences between $S$ and $S'$ are typically minimal if the instance is large enough and the SVC maximum capacity is limited.

For this reason, to avoid complete solution copies, in FILO2 we adopt a technique based on do/undo-lists. This technique assumes that before starting the core optimization procedure, the above mentioned three solutions $S$, $S'$ and $S^*$ are identical. 
In particular, we have that $S^*$ is the solution obtained by the construction phase and $S \gets S' \gets S^*$. This condition can be obtained by performing two full solution copies once only.

At the beginning of every core optimization iteration, an invariant states that $S$ is equal to $S'$. 
Next, shaking and local search applications are applied to $S'$ to obtain a solution which is an actual neighbor of $S$. 
During these applications, every change applied to $S'$ is recorded into a do-list $\mathcal{D}$, and likewise, the inverse change is recorded into an undo-list $\mathcal{U}$. List $\mathcal{D}$ thus contains all changes that should be applied to $S$ to obtain $S'$, and similarly list $\mathcal{U}$ allows to obtain $S$ from $S'$ by applying the stored changes in reverse order. 

Whenever $S'$ is accepted to replace $S$ because of the SA criterion, all changes stored in $\mathcal{D}$ are applied in chronological order to $S$. 
Moreover, these changes are also inserted into an additional do-list $\mathcal{D}^*$, that will possibly at some point be used to convert the best solution $S^*$ into $S'$.
If, on the other hand, $S'$ is not accepted to replace $S$, the undo-list $\mathcal{U}$ is applied to $S'$ in reverse chronological order.
Finally, in both cases, lists $\mathcal{D}$ and $\mathcal{U}$ are cleared.

To conclude, whenever $S'$ is improving with respect to the best solution $S^*$, and thus $S'$ should replace $S^*$, changes in the do-list $\mathcal{D}^*$ are applied to $S^*$, and then $\mathcal{D}^*$ is cleared.

We shall note that converting $S$ into $S'$, through the application of $\mathcal{D}$ to $S$ when the SA acceptance criterion is satisfied, can in practice be avoided since $S$ can always be recovered from the application of $\mathcal{U}$ to $S'$. This makes the existence of $S$ redundant, and in practice not needed. Pseudocode for the updated FILO2 core optimization procedure is shown in Algorithm \ref{algorithm:filo2-coreopt}.
\begin{algorithm}
	\footnotesize
	\caption{FILO2 core optimization procedure}
	\label{algorithm:filo2-coreopt}
	\algcomment{Input parameters: solution $S^*$, random engine $\mathcal{R}$}
	\begin{algorithmic}[1]
		\Procedure{CoreOpt}{$S^*, \mathcal{R}$}
        \State $S' \gets S^*, \mathcal{T} \gets \mathcal{T}_0, c_{S} \gets \Call{Cost}{S'}$
		\State $\bm{\omega} \gets (\omega_1, \omega_2, \ldots, \omega_{|V_c|}), \omega_i \gets \omega_{base} \; \forall i \in V_c$ 
		\State $\bm{\gamma} \gets (\gamma_0, \gamma_1, \ldots, \gamma_{|V_c|}), \gamma_i \gets \gamma_{base} \; \forall i \in V$ 
		\State $\mathcal{D} \gets \mathcal{U} \gets \mathcal{D}^* \gets [ ]$
		\For{$n \gets 1 \text{ to } \Delta_{CO}$}
			\State $S' \gets \Call{Shaking}{S', \mathcal{R}, \bm{\omega}, \mathcal{D}, \mathcal{U}}$
            \State $S' \gets \Call{LocalSearch}{S', \mathcal{R}, \mathcal{D}, \mathcal{U}}$
            \If{$\Call{Cost}{S'} < \Call{Cost}{S^*}$}
                \State $S^* \gets \Call{ApplyDoList}{\mathcal{D}^*, S^*}$
				\State $\mathcal{D}^* \gets []$

                \State $\Call{ResetSparsificationFactors}{\bm{\gamma}}$

            \Else
                \State $\Call{UpdateSparsificationFactors}{\bm{\gamma}}$
            \EndIf
            \State $\Call{UpdateShakingParameters}{\bm{\omega}, S', c_{S}, \mathcal{R}}$
            \If{$\Call{AcceptNeighbor}{c_{S}, S', \mathcal{T}}$}
				\State $c_{S} \gets \Call{Cost}{S'}$
				\State $\mathcal{D}^* \gets \mathcal{D}^* + \mathcal{D}$
			\Else
				\State $S' \gets \Call{ApplyUndoList}{\mathcal{U}, S'}$
			\EndIf
			\State $\mathcal{U} \gets \mathcal{D} \gets []$
            \State $\mathcal{T} \gets c \cdot \mathcal{T}$
		\EndFor

		\State \Return $S^{*}$
		\EndProcedure
	\end{algorithmic}
\end{algorithm}

A similar approach is used in the route minimization procedure to keep the current working solution and the final solution synchronized during the algorithm execution.

In practical terms, the primitive changes supported by this technique are the solution operations happening during the shaking and local search steps, i.e., vertex removal and insertion, empty route creation and removal, one-customer route creation and removal, and the reverse of a contiguous sequence of vertices in a route (required by the intra-route 2-opt local search operator).

By using do/undo-lists, the synchronization between solutions is no longer a linear operation in the instance size. However, it is still bounded by the actual number of changes performed by the shaking and the local search applications which is limited by the SVC and mostly instance size independent.

\subsubsection{Lazy Preprocessing for \textsc{tail} and \textsc{split} Local Search Operators} \label{sec:lazy-tail-split}
The local search engine of FILO employs, among others, two inter-route adaptations of the 2-opt operator (see \citet{Reinelt1994TheTS}) called \textsc{tail} and \textsc{split}, both working on two different routes at a time. 

By denoting as head, a path of vertices belonging to the initial part of a route, and as tail, a path of vertices belonging to the final part of a route, \textsc{tail} swaps the tails of the two involved routes at some point, whereas \textsc{split} cuts the two routes at some point, then it replaces the tail of the first route with the reversed head of the second route and the head of the second route with the reversed tail of the first route.
The execution of such moves, and in particular the check of their feasibility, benefits from the precomputation, at the beginning of the neighborhood exploration of the cumulative load up to $q_{i}^{up}$ and from $q_{i}^{from}$ any customer $i \in V_c$ included, i.e., $q_{i}^{up} = q_i + q_{\pi^1_i} + q_{\pi^2_i} + \ldots + q_0$ and $q_{i}^{from} = q_i + q_{\sigma^1_i} + q_{\sigma^2_i} + \ldots + q_0$ where $\pi^n_i$ and $\sigma^n_i$ identifies the $n$-th predecessor and successor of $i$, respectively.

Knowing $q_{i}^{up}$ and $q_{i}^{from}$ for any customer $i \in V_c$, allows to perform feasibility checks for \textsc{tail} and \textsc{split} in constant-time. Therefore, in FILO these values are precomputed for all customers $i \in V_c$ at the beginning of both the \textsc{tail} and \textsc{split} neighborhood explorations. However, since local search applications are very limited in their scope, precomputing these values for all possible customers is unlikely to be useful and most certainly very time consuming for XXL instances. 
As a consequence, in FILO2, $q_{i}^{up}$ and $q_{i}^{from}$ for customers $i$ belonging to route $r$ are computed on-demand only when required, and cached until route $r$ is not changed. Whenever some edit is performed to $r$, $q_{i}^{up}$ and $q_{i}^{from}$ for $i \in r$ are marked as no longer valid.

\subsubsection{Hierarchical Randomized Variable Neighborhood Descent} \label{sec:hrvnd}
In the original FILO algorithm, local search operators are organized according to the Hierarchical Randomized Variable Neighborhood Descent (HRVND) principle. More precisely, operators of the same cardinality are grouped together in tiers. Each tier can be seen as a compound operator in which operators are applied according to the Randomized Variable Neighborhood (RVND) principle, that is the order of application of the operators is randomly chosen before each tier application and whenever after a complete application of all the operators an improvement is found, all the operators are re-considered (inner RVND loop), see Algorithm \ref{alg:hrvnd-tier} (left). Tiers are then linked together in increasing computational complexity according to the VND principle. Again, whenever after all tiers application, an improvement is found, the complete local search is repeated (outer VND loop).
\begin{algorithm}
	\footnotesize
	\caption{FILO (left) and FILO2 (right) HRVND tier application}\label{alg:hrvnd-tier}
	\algcomment{Notation: $\mathcal{O}$ list of tier operators, $\mathcal{O}_c$ operator in position $c$, $\mathcal{R}$ random engine.}
	\begin{minipage}[t]{0.46\textwidth}
		\begin{algorithmic}[1]
			\Procedure{TierApplication}{$S, \mathcal{O}, \mathcal{R}$}
			\State $\mathcal{O} \gets \Call{Shuffle}{\mathcal{O}, \mathcal{R}}$
			\State $e \gets 0, c \gets 0$
			\Repeat
				\State $S' \gets \Call{Apply}{\mathcal{O}_c, S}$
				\IfThen{$\Call{Cost}{S'} < \Call{Cost}{S}$}{$S \gets S', e \gets c$}
				\State $c \gets (c + 1)\Mod{\Call{Length}{\mathcal{O}}}$
			\Until{$c \neq e$}
			\State \Return $S$
			\EndProcedure
		\end{algorithmic}
	\end{minipage}
	\hfill
	\begin{minipage}[t]{0.46\textwidth}
		\begin{algorithmic}[1]
			\Procedure{TierApplication}{$S, \mathcal{O}, \mathcal{R}$}
			\State $\mathcal{O} \gets \Call{Shuffle}{\mathcal{O}, \mathcal{R}}$
			\For{$o \in \mathcal{O}$}
			\State $S \gets \Call{Apply}{o, S}$
			\EndFor
			\State \Return $S$
			\EndProcedure
		\end{algorithmic}
	\end{minipage}
\end{algorithm}

Novel extensive experimental tests performed with FILO2 have shown that the inner RVND loop is seldom able to bring significant improvements given that the outer VND loop is already performed. In FILO2 we thus simplified the HRVND by removing the inner RVND loop, that is, operators within each tier are executed only once per outer VND loop, see Algorithm \ref{alg:hrvnd-tier} (right). This allows to save some costly operations without significantly decreasing the quality of final solutions.

\section{Computational Testing} \label{sec:computaional_testing}
The computational testing aims at verifying that the above introduced changes to the original algorithm still allow FILO2 to provide state-of-the-art results on well-known literature benchmark instances as well as they make it applicable to the new dataset of XXL instances.
To accomplish the first goal, we test FILO2 on the $\mathbb{X}$ and $\mathbb{B}$ instances proposed by \citet{UCHOA2017845} and \citet{arnold2019xxl}, respectively. The computational results of FILO2 are compared to what was previously obtained by FILO.
Once assessed that FILO2 provides results comparable to those obtained by FILO, we move on to the real target of this research, that is, showing that FILO2 can easily manage much larger instances. To this end, we test the proposed approach on a novel $\mathbb{I}$ dataset containing XXL instances with up to one million customers.

\subsection{Implementation and Experimental Environment}
The proposed algorithm was implemented in C++ and compiled using g++ 11.2. All the experiments were performed using a single thread on a 64-bit GNU/Linux Ubuntu 22.04 mini computer equipped with an AMD Ryzen 5 PRO 4650GE CPU, running at 3.3 GHz, and 16GB of RAM. 

The source code of FILO2 is available at \href{https://github.com/acco93/filo2}{https://github.com/acco93/filo2}.
In the computations we considered a standard version of FILO2, performing  $10^5$ core optimization iterations, and a longer version, called FILO2 (long), which performs $10^6$ iterations. 
For every problem instance, the algorithms were executed for a symbolic number of ten runs. 

We did similarly for the original algorithm FILO, whose source code is available at \href{https://github.com/acco93/filo}{https://github.com/acco93/filo}. The results of FILO have been obtained by running the source code on the above mentioned computing system.
Furthermore, because of their randomized nature, all the analysis on solution quality and computing time described in the following sections refer to the average behavior of the algorithms.
The best solution values (BKS) for $\mathbb{X}$ and $\mathbb{B}$ instances have been taken from \citet{CVRPLIB} at the time of writing, whereas BKS for the new $\mathbb{I}$ instances are the costs of the best solutions we found during the overall experimentation. Finally, gaps are computed as $100 \cdot (z - \text{BKS}) / \text{BKS}$ where $z$ is the final solution value obtained by the algorithms, the computing time is always reported in seconds, and the tables in the subsequent paragraphs always show the average percentage gap obtained by the algorithm with respect to the BKS (Avg).

\subsection{Parameter Tuning} \label{sec:parameter_tuning}
Algorithm FILO2 inherits all the parameters and their values from FILO (see Table \ref{tab:filo-params} for the main ones). Moreover, FILO2 adds an additional parameter $n_{nn}$, first introduced in Section \ref{sec:neighbor_lists}, that identifies the number of nearest neighbors that are computed, stored and used during the algorithm execution.
Whenever $n_{nn} = |V|$, FILO2 would behave similarly to FILO, however, given the storage and computing time requirements it would require, this is not possible for XXL instances. 

The value of $n_{nn}$ was selected with a simple uniform sampling of reasonable values so as to obtain results of FILO2 comparable with those obtained by FILO on the $\mathbb{X}$ and $\mathbb{B}$ literature instances. At the same time, as it is described in Section \ref{sec:analysis}, it was also set to obtain a good tradeoff between solution quality and computing time on the newly proposed $\mathbb{I}$ instances. As a deliberate design choice, all parameters inherited from FILO are set at their original value to both simplify the tuning process and to keep the resulting FILO2 algorithm behaving as similar as possible to the original one.

\subsection{Testing on \texorpdfstring{$\mathbb{X}$}{X} Instances}
\begin{table}
\footnotesize
\centering
\begin{tabular}{ll rrrrrrrrrrrr}
\toprule
 & && \multicolumn{2}{c}{FILO} && \multicolumn{2}{c}{FILO2} && \multicolumn{2}{c}{FILO (long)} && \multicolumn{2}{c}{FILO2 (long)} \\
\cmidrule{4-5}
\cmidrule{7-8}
\cmidrule{10-11}
\cmidrule{13-14}
Size & Vertices && Avg & $t$ && Avg & $t$ && Avg & $t$ && Avg & $t$ \\
\midrule
S&101 - 247&&      0.18&        78&&      0.17&        75&&      0.08&       827&&      0.08&       807\\
M&251 - 491&&      0.39&        73&&      0.36&        73&&      0.25&       771&&      0.23&       769\\
L&502 - 1001&&      0.53&        75&&      0.50&        82&&      0.32&       763&&      0.29&       831\\
\midrule
All&101 - 1001&&      0.37&        75&&      0.34&        76&&      0.22&       786&&      0.20&       801\\
\bottomrule
\end{tabular}
\caption{Comparison of FILO and FILO2 on $\mathbb{X}$ instances.}
\label{tab:x_aggregates}
\end{table}
The $\mathbb{X}$ dataset proposed by \citet{UCHOA2017845} is the main reference literature dataset for the CVRP containing a hundred medium to large instances with a variety of customer demand distributions and vertex layouts. 
Table \ref{tab:x_aggregates} compares the results of the proposed approach with those of the original one. 

As can be seen from the table, the solution quality obtained by both the standard and longer version of FILO2 is comparable with that obtained by FILO. 
More precisely, we have that FILO and FILO2, as well as their longer version, are not significantly different in the quality of final solutions on the instances with S size. On the other hand, FILO2 always finds statistically better solutions than FILO on instances with M and L sizes (see Appendix \ref{appendix:x_instances} for more details). As far as the computing time is concerned, the new approach behaves in average very similarly to the original one. Interestingly, FILO2 seems to require a longer computing time on L sized instances. However, given the results on much larger instances (see Section \ref{sec:b-results}), we believe this to be mainly related to the slightly different search trajectory of the two algorithms, rather than because of obvious inefficiencies introduced in FILO2.

\subsection{Testing on \texorpdfstring{$\mathbb{B}$}{B} Instances} \label{sec:b-results}
The $\mathbb{B}$ instances proposed by \citet{arnold2019xxl} are a set of ten very large-scale instances containing up to thirty thousand customers and reflecting real-world parcel distribution problems in Belgium.
They include a first scenario, in which the depot is located centrally with respect to the customers and relatively short routes are performed, and a second one in which the depot is eccentric with respect to the service zone, and thus much longer routes are required to visit the customers.
\begin{table}[b]
	\footnotesize
	\centering
	\begin{tabular}{ll rrrrrrrrrrrr}
	\toprule
		& && \multicolumn{2}{c}{FILO} && \multicolumn{2}{c}{FILO2} && \multicolumn{2}{c}{FILO (long)} && \multicolumn{2}{c}{FILO2 (long)} \\
	\cmidrule{4-5}
	\cmidrule{7-8}
	\cmidrule{10-11}
	\cmidrule{13-14}
	ID ($|V_c|$) &  BKS && Avg & $t$ && Avg & $t$ && Avg & $t$ && Avg & $t$ \\
	\midrule
L1 \tiny{(3000)} & 192848 &&       0.53 &         95 &&       0.42 &         98 &&       0.26 &        958 &&       0.20 &        992\\
L2 \tiny{(4000)} & 111391 &&       1.03 &        142 &&       0.85 &        130 &&       0.49 &       1515 &&       0.32 &       1479\\
A1 \tiny{(6000)} & 477277 &&       0.61 &        124 &&       0.52 &        105 &&       0.29 &       1247 &&       0.25 &       1073\\
A2 \tiny{(7000)} & 291350 &&       1.10 &        139 &&       1.08 &        114 &&       0.44 &       1505 &&       0.40 &       1226\\
G1 \tiny{(10000)} & 469531 &&       0.74 &        161 &&       0.66 &        114 &&       0.30 &       1602 &&       0.27 &       1180\\
G2 \tiny{(11000)} & 257748 &&       1.51 &        198 &&       1.47 &        127 &&       0.44 &       2284 &&       0.41 &       1505\\
B1 \tiny{(15000)} & 501719 &&       1.08 &        197 &&       1.02 &        113 &&       0.42 &       2006 &&       0.39 &       1206\\
B2 \tiny{(16000)} & 345481 &&       2.01 &        228 &&       1.91 &        122 &&       0.58 &       2564 &&       0.55 &       1430\\
F1 \tiny{(20000)} & 7240124 &&       0.81 &        303 &&       0.73 &        136 &&       0.35 &       3276 &&       0.31 &       1575\\
F2 \tiny{(30000)} & 4373320 &&       2.11 &        482 &&       2.13 &        153 &&       0.68 &       6194 &&       0.62 &       2040\\
\midrule
Mean &  &&       1.15 &        207 &&       1.08 &        121 &&       0.42 &       2315 &&       0.37 &       1371\\
	\bottomrule
	\end{tabular}
	\caption{Comparison of FILO and FILO2 on the $\mathbb{B}$ dataset.}
	\label{tab:b-computations}
\end{table}
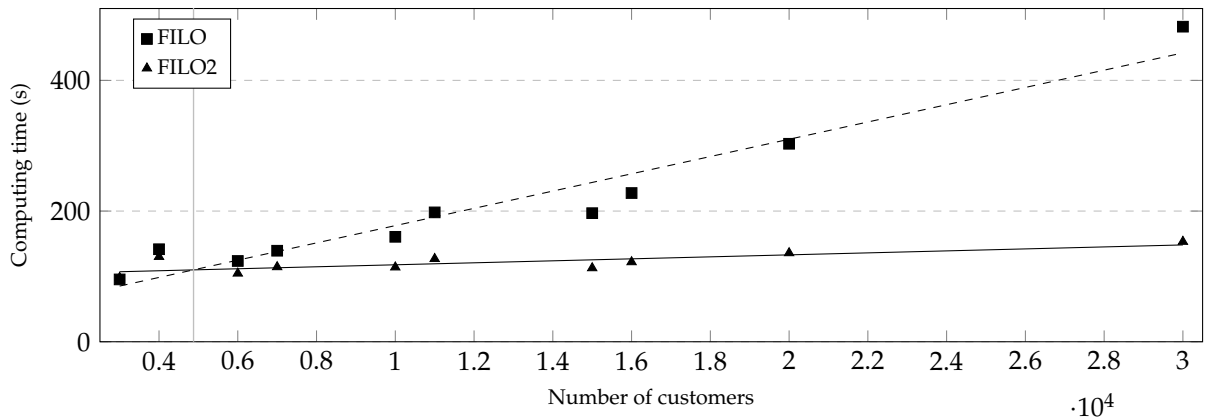
\begin{figure}
    \centering
       \caption{Computing time growth of FILO and FILO2 on the $\mathbb{B}$ dataset. The dashed and the continuous lines show a linear function fitting the existing data for FILO and FILO2, respectively. The two linear functions met when the number of customers is approximately $4876$.}
    \begin{tikzpicture}

\begin{axis}[
    title={},
    ymajorgrids=true,
    grid style=dashed,	
    xlabel={\footnotesize Number of customers},
    ylabel={\footnotesize Computing time (s)},
    width=\textwidth, height=6cm,
    legend pos=north west,
    legend cell align={left},
    ymin=0,
    ymax=510,
    xmin=2500,
    xmax=30500
]

\addplot[only marks, mark=square*, mark size=2pt] coordinates {
(3000, 95.4)(4000, 141.6)(6000, 123.6)(7000, 139.3)(10000, 160.6)(11000, 198)(15000, 196.8)(16000, 227.6)(20000, 303)(30000, 482.2)
};

\addplot[only marks, mark=triangle*, mark size=2pt] coordinates {
(3000, 98.0)(4000, 130.1)(6000, 104.6)(7000, 114.4)(10000, 114.0)(11000, 126.9)(15000, 112.8)(16000, 122.0)(20000, 135.8)(30000, 153.2)
};

\addplot[domain=3000:30000, dashed]{0.01321693*x+45.563406029506126};

\addplot[domain=3000:30000]{0.00152*x+102.57989};

\addplot [mark=none, line width=0.5pt, gray!50] coordinates {(4876.39857694,0) (4876.39857694,600)};

\legend{\footnotesize{FILO}, \footnotesize{FILO2}}
	
\end{axis}

\end{tikzpicture}
\label{fig:timegrowth-b-dataset}
\end{figure}
Table \ref{tab:b-computations} compares FILO and FILO2 configurations and Figure \ref{fig:timegrowth-b-dataset} provides some hints on the computing time growth of the two algorithms. 

The results show that both FILO2 and FILO2 (long) are on average better than the original approach. Moreover, statistical analysis shows that the final solution quality of FILO2 and FILO is not significantly different. On the other hand, the quality of final solutions generated by FILO2 (long) is significantly better than that of solutions generated by FILO (long). For more details see Appendix \ref{appendix:b_instances}. In addition, Figure \ref{fig:timegrowth-b-dataset} shows that the computing time of FILO2 increases at a much slower rate compared to that of FILO. More precisely, it seems that the newly introduced changes are more convenient in terms of computing time on very large-scale instances with at least $5000$ customers, making the overall algorithm considerably more scalable when much larger instances are solved. Finally, from the actual observations, FILO2 results to be slower than FILO on the smallest L1 instance only.

\subsection{Testing on \texorpdfstring{$\mathbb{I}$}{I} Instances}
We propose the new $\mathbb{I}$ dataset containing twenty XXL instances having a number of customers ranging from $20,000$ to $1,000,000$. These instances are obtained by projecting to the Euclidean plane, geographical coordinates of randomly sampled addresses in various Italian regions. The addresses are based on \citet{OpenAddresses} data. Similarly to the $\mathbb{B}$ instances, customers have a load that is uniformly randomly selected in $[1, 3]$. The depot is located on the regional capital city position and the vehicle capacity is $Q=50$ when the depot is located centrally in the region. Otherwise, the capacity is  $Q=150$  when the depot is eccentric, and $Q=200$ when very close to the region boundary, respectively. As a consequence, the dataset contains half of the instances where we can expect relatively short routes, and half where longer routes are required. The dataset can be downloaded from \href{https://github.com/acco93/filo2}{https://github.com/acco93/filo2}. Finally, the new instances follow the same conventions followed by the $\mathbb{X}$ and $\mathbb{B}$ datasets, that is, vertex coordinates are integer and the arc costs are rounded to the nearest integer.
\begin{table}
\footnotesize
\centering
\begin{tabular}{l lr rrrrrrrrrr }
\toprule
 & && && & \multicolumn{2}{c}{FILO2} && \multicolumn{2}{c}{FILO2 (long)} \\
\cmidrule{7-8}
\cmidrule{10-11}
ID ($|V_c|\cdot 10^3$) & Depot & $Q$ & BKS & Routes && Avg & $t$ && Avg & $t$ \\
\midrule
Valle-D-Aosta \tiny{(20)} & C & 50 & 21679514 & 800 &&       0.28 &        192 &&       0.13 &       2620\\
Molise \tiny{(50)} & C & 50 & 111184982 & 2007 &&       0.41 &        146 &&       0.16 &       1901\\
Trentino-Alto-Adige \tiny{(100)} & S & 150 & 102063181 & 1338 &&       0.92 &        194 &&       0.40 &       2292\\
Basilicata \tiny{(150)} & W & 150 & 175623919 & 2007 &&       0.74 &        185 &&       0.31 &       2722\\
Umbria \tiny{(200)} & C & 50 & 545507981 & 8006 &&       0.29 &        248 &&       0.12 &       2124\\
Abruzzo \tiny{(250)} & NW & 200 & 311712556 & 2500 &&       0.73 &        278 &&       0.28 &       2613\\
Friuli-Venezia-Giulia \tiny{(300)} & SE & 200 & 415805616 & 3004 &&       0.60 &        393 &&       0.25 &       4116\\
Liguria \tiny{(320)} & C & 50 & 1426389867 & 12801 &&       0.20 &        332 &&       0.07 &       2964\\
Calabria \tiny{(380)} & C & 50 & 1964651530 & 15201 &&       0.40 &        292 &&       0.13 &       2299\\
Marche \tiny{(420)} & E & 200 & 420484426 & 4205 &&       0.69 &        408 &&       0.26 &       3634\\
Sardegna \tiny{(470)} & S & 200 & 827934149 & 4732 &&       1.61 &        316 &&       0.78 &       2690\\
Campania \tiny{(500)} & SW & 200 & 391859276 & 5003 &&       1.01 &        420 &&       0.37 &       4417\\
Piemonte \tiny{(600)} & C & 50 & 2627446164 & 23992 &&       0.32 &        404 &&       0.15 &       2721\\
Toscana \tiny{(700)} & N & 150 & 1084417188 & 9336 &&       0.77 &        488 &&       0.33 &       5793\\
Puglia \tiny{(750)} & E & 200 & 1464797603 & 7538 &&       1.49 &        436 &&       0.74 &       2957\\
Sicilia \tiny{(800)} & NW & 200 & 1774262462 & 8044 &&       1.35 &        469 &&       0.63 &       3298\\
Veneto \tiny{(850)} & E & 200 & 1050488613 & 8499 &&       0.71 &        600 &&       0.29 &       5131\\
Emilia-Romagna \tiny{(900)} & C & 50 & 5405446715 & 36001 &&       0.24 &        501 &&       0.09 &       3263\\
Lombardia \tiny{(950)} & SW & 150 & 1339900081 & 12669 &&       0.75 &        562 &&       0.34 &       7990\\
Lazio \tiny{(1000)} & C & 50 & 3145381332 & 39982 &&       0.40 &        532 &&       0.14 &       3938\\
\midrule
Mean &  &&&&&       0.70 &        370 &&       0.30 &       3474\\
\bottomrule
\end{tabular}
\caption{Detailed results obtained by FILO2 on the new $\mathbb{I}$ instances.}
\label{tab:i-instances}
\end{table}

Computational results of FILO2 and some instance statistics are shown in Table \ref{tab:i-instances}. In particular, for each instance we report its name and size in thousands of customers, the depot location in cardinal points with the addition of C to denote a central depot, and the vehicle capacity $Q$.
Furthermore, the table reports also the cost and number of routes of the best known solution obtained during our entire experimentation as well as the average gap and computing time in seconds for both FILO2 and FILO2 (long). Detailed results are available in Appendix \ref{appendix:i_instances}.

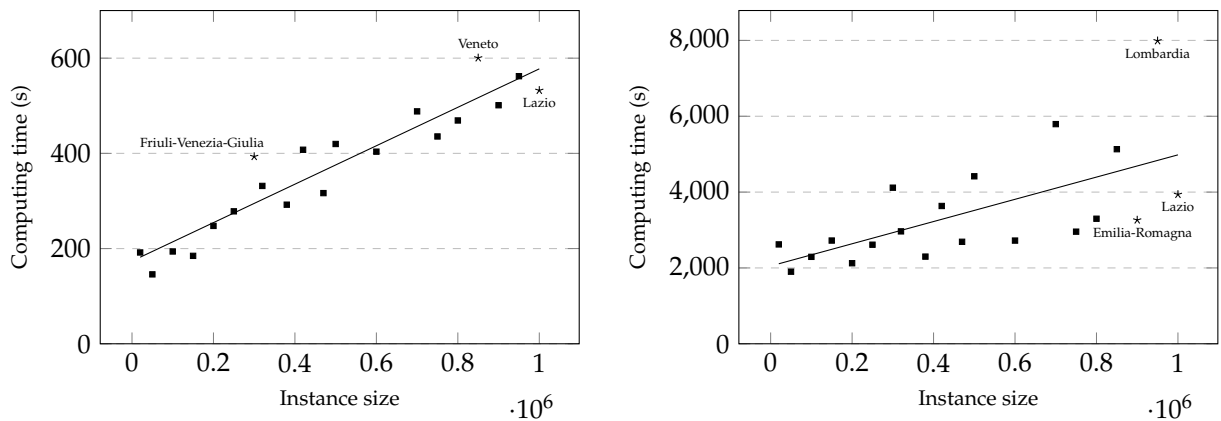
\begin{figure}
    \centering
    \caption{Computing time growth of FILO2 (left) and FILO2 (long) (right) on the $\mathbb{I}$ dataset. Marks represent the algorithms computing time, while the line linearly fits the data. The most three most influential observations according to the Cook's distance are marked with a star.}
    \begin{tikzpicture}

\begin{axis}[
    title={},
    ymajorgrids=true,
    grid style=dashed,	
    xlabel={\footnotesize Instance size},
    ylabel={\footnotesize Computing time (s)},
    width=0.49 * \textwidth, height=6cm,
    ymin=0,
    ymax=700,
]

\addplot[only marks, mark=square*, mark size=1pt] coordinates {
(20000, 191.8)(50000, 145.8)(100000, 193.9)(150000, 184.7)(200000, 247.6)(250000, 278.2)(320000, 331.6)(380000, 292.2)(420000, 407.6)(470000, 316.5)(500000, 419.7)(600000, 403.6)(700000, 488.2)(750000, 435.5)(800000, 469.1)(900000, 501.1)(950000, 562.1)
};

\addplot[only marks, mark=star, mark size=1.6pt, point meta=explicit symbolic, nodes near coords, text width=3cm, align=left, font=\tiny\linespread{0.8}\selectfont] coordinates {
(300000, 393.3)[Friuli-Venezia-Giulia]
};

\addplot[only marks, mark=star, mark size=1.6pt, point meta=explicit symbolic, nodes near coords, text width=3cm, align=center, font=\tiny\linespread{0.8}\selectfont] coordinates {
(850000, 600.5)[Veneto]
};

\addplot[only marks, mark=star, mark size=1.6pt, point meta=explicit symbolic, nodes near coords, text width=1cm, align=center, every node near coord/.append style={yshift=-0.35cm}, font=\tiny\linespread{0.8}\selectfont] coordinates {
    (1000000, 532.3)[Lazio]
};

\addplot[domain=20000:1000000]{0.0004038014398813907*x+173.71940093758474};

\end{axis}

\end{tikzpicture}
\hfill
\begin{tikzpicture}

    \begin{axis}[
        title={},
        ymajorgrids=true,
        grid style=dashed,	
        xlabel={\footnotesize Instance size},
        ylabel={\footnotesize Computing time (s)},
        width=0.49 * \textwidth, height=6cm,
        ymin=0,
    ]
    
    \addplot[only marks, mark=square*, mark size=1pt] coordinates {
        (20000, 2620.4)(50000, 1901.3)(100000, 2292.5)(150000, 2722.2)(200000, 2123.7)(250000, 2612.7)(300000, 4116.3)(320000, 2964.2)(380000, 2298.7)(420000, 3633.8)(470000, 2689.7)(500000, 4416.7)(600000, 2721.3)(700000, 5792.7)(750000, 2956.8)(800000, 3297.7)(850000, 5131.4)
    };

    \addplot[only marks, mark=star, mark size=1.6pt, point meta=explicit symbolic, nodes near coords, text width=1cm, align=center, every node near coord/.append style={yshift=-0.35cm}, font=\tiny\linespread{0.8}\selectfont] coordinates {
    (950000, 7989.5)[Lombardia]
    (1000000, 3938.1)[Lazio]
    };

    \addplot[only marks, mark=star, mark size=1.6pt, point meta=explicit symbolic, nodes near coords, text width=1.45cm, align=right, every node near coord/.append style={yshift=-0.40cm}, font=\tiny\linespread{0.8}\selectfont] coordinates {
    (900000, 3263.1)[Emilia-Romagna]
    };

    \addplot[domain=20000:1000000]{0.0029308095045889977*x+2051.231985522041};

    \end{axis}
    
    \end{tikzpicture}
    \label{fig:timegrowth-i-dataset}
\end{figure}

To better determine the behavior of the approach, Figure \ref{fig:timegrowth-i-dataset} shows the computing time growth of the algorithm configurations on this new dataset. As can be seen from the charts both variants exhibit a linear growth of their computing time. On the one hand, FILO2 seems to generate results with generally less variance of the computing time with respect to the predicted value given by the linear regression on the observed data. 
On the other hand, the distance of points from the regression line seems to increase with the instance size for FILO2 (long).
Highly influential points, i.e., points that if removed would considerably change the regreession line, can be identified by computing the Cooks's distance (see \citet{cooksdistance}). The three most influential points, in decreasing order of influence, for both algorithms, are Veneto (0.247811), Friuli-Venezia-Giulia (0.174067), Lazio (0.135176) for FILO2 and Lombardia (0.850441), Emilia-Romagna (0.141266), and Lazio (0.114841) for FILO2 (long). 

It is difficult to determine the reasons for the abnormally larger computing time of FILO2 (long) on instance Lombardia. We can however observe that the running time of the standard FILO2 version on such instance is in line with those of the other instances of similar size.

An additional interpretation could be that it is not instance Lombardia requiring a long computing time when solved by FILO2 (long) but rather that instances with similar sizes are indeed solved faster by FILO2 (long) compared to what we could expect by a tenfold increase in the number of iterations of FILO2. In fact, contrarily to what is happening for the $\mathbb{X}$ and $\mathbb{B}$ datasets, the average computing time of FILO2 (long) on the $\mathbb{I}$ instances is smaller than ten times the computing time of FILO2. More specifically, this is happening for 12 instances out of 20 sometimes with unexpectably low values (e.g., see instance Emilia-Romagna).

\section{Algorithmic Components Analysis} \label{sec:analysis}
This section provides some insights on the effectiveness of the newly introduced changes in FILO2. If not stated otherwise, the analysis are performed on the $\mathbb{I}$ instances by running the the standard FILO2 configuration for $10^5$ core optimization iterations. Computational results always refer to the average of ten distinct runs.

\subsection{Number of Neighbors} \label{sec:analysis_instance_preprocessing}
When moving to XXL instances, the initial computation of the limited neighbor lists $\mathcal{N}^{n_{nn}}_i(V), i \in V$ may take a significant portion of the overall algorithm execution time, especially when relatively few core optimization iterations are performed, as in the standard FILO2 version. 
As an example, for the largest instance (Lazio), the initial preprocessing performed by FILO2 takes approximately  $48\%$ of the total computing time. Table \ref{tab:i-preprocessing-time} shows, for every instance in the $\mathbb{I}$ dataset, the running time spent to compute the neighbor lists and the percentage of the overall algorithm computing time for both FILO2 and FILO2 (long).

Luckily this procedure could in practice be easily parallelized, thus the process could be sped up by employing more CPU cores at the same time. Finally, in scenarios in which parallelization is not allowed but initial solutions are required to be generated as fast as possible, e.g., the \citet{DIMACS} competition where algorithm were evaluated according to the primal integral metric proposed in \citet{Achterberg2012}, one could employ more sophisticated approaches based on a lazy computation of neighbors for a vertex as they are required by the algorithm procedures, e.g., initially only $n_{cw} < n_{nn}$ neighbors are computed so that they can be used during the initial construction phase based on the limited savings algorithm (see Section \ref{sec:filo}), then the list of neighbors for specific vertices $i$ is enlarged as soon as it is required, e.g., when a recreate application tries to reinsert a customer $i$ back into the solution. 
\begin{table}
	\footnotesize
	\centering
	\begin{tabular}{l rrr }
	\toprule
	ID & $t (s)$ & FILO2 (\%) & FILO2 (long) (\%) \\
	\midrule
	Valle-D-Aosta &       4.00 &       2.09 &       0.15 \\
Molise &      10.00 &       6.86 &       0.53 \\
Trentino-Alto-Adige &      23.00 &      11.86 &       1.00 \\
Basilicata &      33.00 &      17.87 &       1.21 \\
Umbria &      49.00 &      19.79 &       2.31 \\
Abruzzo &      61.00 &      21.93 &       2.33 \\
Friuli-Venezia-Giulia &      75.00 &      19.07 &       1.82 \\
Liguria &      78.00 &      23.52 &       2.63 \\
Calabria &      92.00 &      31.49 &       4.00 \\
Marche &     103.00 &      25.27 &       2.83 \\
Sardegna &     113.10 &      35.73 &       4.20 \\
Campania &     127.00 &      30.26 &       2.88 \\
Piemonte &     153.00 &      37.91 &       5.62 \\
Toscana &     177.60 &      36.38 &       3.07 \\
Puglia &     178.10 &      40.90 &       6.02 \\
Sicilia &     197.90 &      42.19 &       6.00 \\
Veneto &     222.20 &      37.00 &       4.33 \\
Emilia-Romagna &     234.60 &      46.82 &       7.19 \\
Lombardia &     241.90 &      43.04 &       3.03 \\
Lazio &     258.60 &      48.58 &       6.57 \\
\midrule
Mean &     121.60 &      32.89 &       3.50 \\
	\bottomrule
	\end{tabular}
	\caption{Preprocessing time used to compute the neighbor lists and the percentage of the total computing time of FILO2 and FILO2 (long) spent performing this initial instance preprocessing.}
	\label{tab:i-preprocessing-time}
\end{table}

As mentioned in Section \ref{sec:parameter_tuning}, the computing time for the definition of neighbor lists is affected by parameter $n_{nn}$. This not only affects the initial preprocessing time but also the overall algorithm computing time, the memory requirements as well as the final solution quality.
\begin{figure}
	\centering
	\caption{FILO2 average computing time and final solution quality on the $\mathbb{I}$ dataset while varying parameter $n_{nn}$.}
	\label{fig:nn-tuning}
			\begin{tikzpicture}[scale=0.91]
	\begin{axis}[
	title={},
	ymajorgrids=true,
	grid style=dashed,
	black, fill=black,	
	xlabel={\footnotesize Average computing time (sec)},
	ylabel={\footnotesize Average \% gap},
	enlarge x limits=0.13
	]

\addplot[only marks, mark=square*, mark size=1pt, point meta=explicit symbolic, nodes near coords, text width=3cm, align=center, font=\tiny\linespread{0.8}\selectfont] coordinates {

(169.66, 2.22)[250]
(211.24, 1.53)[500]
(250.91, 1.16)[750]
(290.7, 0.94)[1000]
(330.86, 0.78)[1250]
(369.77, 0.7)[1500]
(413.72, 0.63)[1750]
(461.08, 0.56)[2000]
(511.79, 0.52)[2250]
(562.75, 0.48)[2500]

};

	\end{axis}
	\end{tikzpicture}
\end{figure}
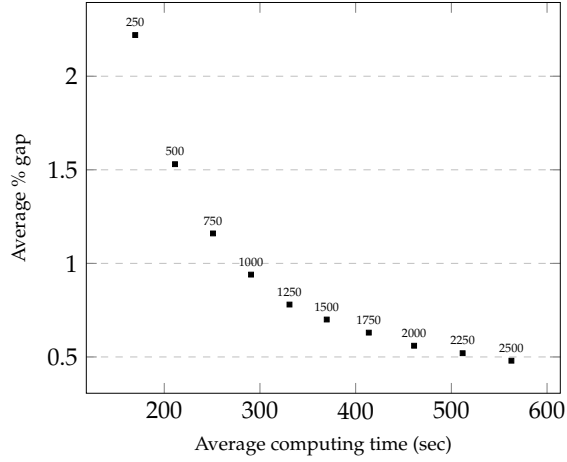
Figure \ref{fig:nn-tuning} shows the overall computing time as well as the final solution quality for several value of $n_{nn}$. Values larger than 2500 would require, for the largest $\mathbb{I}$ instances, more memory than that available in the computing system used for experimentation.

As can be seen from the chart, most of the quality improvement happens for $n_{nn}$ lower than 1250, then the improvement ratio decreases even though it still remains improving while the number of neighbors is increased. The selected parameter $n_{nn} = 1500$ provides a satisfying tradeoff between computing time and final solution quality.

\subsubsection{Approximated Simulated Annealing Temperature}
The initial and final SA temperatures $\mathcal{T}_0$ and $\mathcal{T}_f$ are defined to be proportional to the average arc cost $d$ of the instance under examination, and more precisely $\mathcal{T}_0 = 0.1 \cdot d$ and $\mathcal{T}_f = 0.01 \cdot \mathcal{T}_0$. As mentioned in Section \ref{sec:arc-cost-sampling}, computing $d$ may be too time consuming for XXL instances. In FILO2 we thus use an approximation $\tilde{d}$ defined as the average arc cost obtained by randomly sampling $|V|$ instance arcs. In this section, we compare the quality and computing time of this approximation with respect to the exact computation.

For the $\mathbb{X}$ and $\mathbb{B}$ datasets, the time to compute $d$ is negligible, and in particular, it is about 3 seconds for the largest F2 instance. The value of $\mathcal{T}_0$ is $47.50$ and $186.45$ for the $\mathbb{X}$ and $\mathbb{B}$ datasets averaged over all instances when computing with $d$, and on average $47.27$ and $186.46$ when computed with $\tilde{d}$.

\begin{table}
	\footnotesize
	\centering
	\begin{tabular}{l rrrrr}
	\toprule
	& \multicolumn{2}{c}{Exact ($\mathcal{T}_0 = 0.1 \cdot d$)} && \multicolumn{2}{c}{Approximated ($\mathcal{T}_0 = 0.1 \cdot \tilde{d}$)} \\
	\cmidrule{2-3}
	\cmidrule{5-6}
	ID & $\mathcal{T}_0$ & $t (s)$ && $\mathcal{T}_0$ & $t (ms)$ \\ 
	\midrule
	Valle-D-Aosta &    1784.73 &       1.32 &&    1780.85 &       0.00 \\
	Molise &    3558.74 &       8.27 &&    3553.99 &       0.00 \\
	Trentino-Alto-Adige &    4809.19 &      33.11 &&    4810.42 &       1.20 \\
	Basilicata &    5586.32 &      74.51 &&    5588.64 &       3.20 \\
	Umbria &    4354.79 &     132.48 &&    4355.77 &       6.00 \\
	Abruzzo &    5452.96 &     207.02 &&    5456.59 &      11.00 \\
	Friuli-Venezia-Giulia &    4463.20 &     298.18 &&    4462.78 &      15.10 \\
	Liguria &    7347.70 &     339.24 &&    7346.42 &      16.90 \\
	Calabria &    8155.82 &     478.45 &&    8156.91 &      21.70 \\
	Marche &    5599.47 &     584.64 &&    5602.36 &      24.20 \\
	Sardegna &    8681.42 &     732.39 &&    8683.78 &      28.40 \\
	Campania &    5047.63 &     828.64 &&    5049.10 &      30.40 \\
	Piemonte &    7377.79 &    1193.74 &&    7375.64 &      37.90 \\
	Toscana &    6825.34 &    1624.52 &&    6823.56 &      47.20 \\
	Puglia &   11562.96 &    1864.79 &&   11561.90 &      48.70 \\
	Sicilia &   11035.42 &    2122.36 &&   11038.61 &      52.40 \\
	Veneto &    6388.52 &    2395.48 &&    6389.44 &      55.60 \\
	Emilia-Romagna &    8527.99 &    2686.10 &&    8526.11 &      59.20 \\
	Lombardia &    6767.92 &    2993.95 &&    6768.97 &      65.50 \\
	Lazio &    5711.48 &    3315.50 &&    5709.98 &      65.70 \\
	\midrule
	Mean &    6451.97 &    1095.73 &&     6452.09 &     29.51 \\
	\bottomrule
	\end{tabular}
	\caption{Comparison of exact and approximate SA initial temperature $\mathcal{T}_0$. Note that for the exact approach the computing time is in seconds, while for the approximated one the time is given in milliseconds.}
	\label{tab:sa-comparison}
	\end{table}
The difference in computing time increases considerably for the $\mathbb{I}$ instances. More precisely, the average running time for computing $d$ is more than 1000 seconds while $\tilde{d}$ is computed in less than $30$ milliseconds. The value of $\mathcal{T}_0$ is $6451.97$ when computed with $d$, and on average $6452.09$ when computed with $\tilde{d}$. Table \ref{tab:sa-comparison} provides more details.

Finally, because of its heuristic usage, we consider $\tilde{d}$ to be a good estimation of $d$ with an error of about 0.001\%, and we believe the two values to be close enough for FILO2 to explore values of SA temperatures similar to those explored by FILO.

\subsubsection{Accessing Arc Costs} \label{sec:analysis-accesing-arc-costs}
Given that it is widely used in most of the algorithm procedures, retrieving arc costs is a critical operation that if not properly handled may become one of the main algorithm bottleneck. As described in Section \ref{sec:cost-matrix}, memory constraints impose that the cost matrix cannot be explicitly stored, but costs need to be computed on-demand or retrieved from a limited-size cache. 
\begin{table}[b]
\footnotesize
\centering
\begin{tabular}{l r}
\toprule
 Configuration & $t$  \\
\midrule
Cached   &       471 \\
Cached$^{+}$ &       419 \\
On-demand&       409 \\
On-demand$^{+}$&       370 \\
\bottomrule
\end{tabular}
\caption{Average FILO2 computing time for different cost retrieval strategies. \\
$^{+}$Some costs are retrieved in constant-time from the current solution and from move generators.}
\label{tab:cost_access_comparison}
\end{table}
Table \ref{tab:cost_access_comparison} compares the computing time of FILO2 when different approaches are used to handle cost retrieval. In particular, it compares the algorithm performance when costs are computed on-demand or cached as described in \citet{bentley1990k}. Moreover, for each possibility we consider the effect of using $O(|V| + |T|)$ additional space, where $V$ is the set of vertices and $T$ is the set of move generators, to store frequently used costs in the solution and in move generators to allow for some constant-time cost retrieval. In the table these versions are identified by the $^+$ symbol.

As one would expect, the $^+$ versions using additional space to allow some constant-time retrieval of costs are always preferable compared to the respective version not exploiting this possibility. More surprisingly, the versions in which costs are computed on-demand are considerably more efficient than the versions using an additional caching mechanism.

The cache implementation follows the suggestions described in \citet{bentley1990k}. In particular, we implemented a simple linear cache composed of a signature and a value vectors, respectively $sig$ and $val$, having a size equal to the smallest power of two greater than or equal to the instance size $|V|$. 

Given an arc $(i, j)$ normalized such that $i \leq j$, first a hash index $h$ is computed as $h = i \oplus j$ where $\oplus$ represents a bitwise xor operation. 
The caching algorithm then checks the signature entry $sig[h]$ and if $sig[h] \not= i$ it recomputes the distance $c_{ij}$, sets $val[h] = c_{ij}$, and updates $sig[h] = i$. On the other hand if $sig[h] = i$, it immediately returns $val[h]$ which will contain the correct cost of arc $(i, j)$. This works since $sig[h]$ represents a signature of the arc $(i, j)$ given that $h = i \oplus j$, thus $j = h \oplus i$.
\begin{table}
	\footnotesize
	\centering
	\caption{Average cache percentage hit ratio.}
	\label{tab:hit-ratio}
	\begin{tabular}{l r r}
	\toprule
	ID & Cached & Cached$^{+}$ \\
	\midrule
Valle-D-Aosta &      91.24  &      84.37 \\
Molise &      93.18  &      87.34 \\
Trentino-Alto-Adige &      90.17  &      82.34 \\
Basilicata &      89.77  &      81.65 \\
Umbria &      93.74  &      88.20 \\
Abruzzo &      89.62  &      81.49 \\
Friuli-Venezia-Giulia &      89.64  &      81.44 \\
Liguria &      93.46  &      87.71 \\
Calabria &      93.75  &      88.19 \\
Marche &      89.89  &      81.88 \\
Sardegna &      87.94  &      78.83 \\
Campania &      89.03  &      80.54 \\
Piemonte &      93.62  &      88.01 \\
Toscana &      90.52  &      82.92 \\
Puglia &      89.15  &      80.72 \\
Sicilia &      89.00  &      80.51 \\
Veneto &      89.75  &      81.74 \\
Emilia-Romagna &      93.28  &      87.46 \\
Lombardia &      89.59  &      81.53 \\
Lazio &      92.97  &      86.93 \\
\midrule
Mean &      90.99 &      83.67\\
	\bottomrule
	\end{tabular}
\end{table}

In \citet{bentley1990k}, the author states that the cache will prove effective if the underlying algorithm displays a great locality in the computation of the cost values. Given its optimization pattern, FILO2 should definitely have this locality property. In fact, Table \ref{tab:hit-ratio} shows that the average cache hit ratio is considerably high for both versions using cached costs. The better performance of the on-demand versions might thus be related to hardware advancements rather than to a low usage of the cache. As a final note, we specify that, despite all datasets used in this study consider integer arc costs, in FILO2 arc costs are stored and used as floating point numbers with double precision.

\subsubsection{Recreate Strategy}
Part of the recreate strategy tuning is described in Section \ref{sec:analysis_instance_preprocessing} where the final solution quality and computing time is analyzed while varying parameter $n_{nn}$ that is also affecting the breadth of the recreate procedure as described in Section \ref{sec:recreate}. 

In this section, we analyze what happens if the recreate of the original FILO algorithm, which greedily reinserts a removed customer in its best possible position considering the whole solution, is employed to solve the new $\mathbb{I}$ instances. 

The total average computing time increases to $1224$ seconds compared to the $370$ seconds of the restricted approach. Moreover, the average solution quality decreases to $1.02\%$ compared to $0.70\%$ of the current FILO2 approach. This experimentally shows that carefully restricting reinsertions to close points for large enough instances does not necessarily harm the final solution quality but, on the other hand, may positively influence the search trajectory.

\subsubsection{Solution Copies}
The larger the instance the more convenient becomes the synchronization between solutions by using the do/undo-lists described in Section \ref{sec:do-undo-lists}. 

The average computing time when performing standard solution copies is 2011 seconds, that is more than 5 times larger than the computing time of the algorithms when using the do/undo-lists approach. 

\subsubsection{Lazy Updates for \textsc{tail} and \textsc{split} Local Search Operators}
It's easy to see that precomputing all $q^{up}_i$ and $q^{from}_i$ for every customer $i \in V_c$ as it is described in Section \ref{sec:lazy-tail-split} cannot be beneficial in any way for very large instances. It is in fact incredily expensive since these computations are performed whenever a \textsc{tail} or \textsc{split} neighborhood is explored (possibly several times in a VND loop). In practice, the computing time of the algorithm when these values are always precomputed for all customers is 15843 seconds, i.e., more than 40 times more expensive than the version using lazy updates. 

\subsubsection{Hierarchical Randomized Variable Neighborhood Descent}
The average gap of FILO2 with the standard RVND loop (see Section \ref{sec:hrvnd}) is 0.69\% while when only a single scan of the operators is performed the average gap is 0.70\%. On the other hand, the average computing time of the first version is 413 seconds, while the computing time of the second one is 370. Finally, if we analyze the performance of FILO2 (long) we have that for both versions the average gap is 0.30\% but the computing time for the version with the standard RVND loop is 4085 versus 3474 for the single-scan one. We can conclude that looping through the operators multiple times provides some limited quality advantage on short runs but it does not result on average beneficial on longer runs.

\section{Conclusions}
This paper introduces the $\mathbb{I}$ dataset for CVRP containing twenty XXL instances with a number of customers ranging from 20,000 to 1,000,000. 
The design of extremely efficient algorithms is crucial when dealing with such extremely large instance sizes which are intractable for most existing algorithms for CVRP. We believe these new instances may foster fresh research interest in this direction. The $\mathbb{I}$ instances are solved with FILO2, an evolution of the FILO algorithm proposed in \citet{filo}. FILO2 is able to scale to these sizes still keeping, and sometimes improving, the performance of the original approach on existing well-known literature instances. Computational results are obtained with an ordinary computing system in a considerably short computing time, showing the effectiveness of the proposed changes when combined with the techniques already introduced in \citet{filo}.

\section{Acknowledgments}
We are kindly grateful to Francesco Cavaliere for several discussions about this work and for his contribution to part of the source code that was used as a base for the proposed approach.
The research of Daniele Vigo is partially supported by the U.S. Air Force Office of Scientific Research [Award FA8655-21-1-7046].

\clearpage

\appendix

\section{Instances Layout}
\begin{figure}[!b]
	\captionsetup[subfigure]{labelformat=empty}
	\footnotesize
	\centering
     \begin{subfigure}[b]{0.33\textwidth}
         \includegraphics[width=\textwidth]{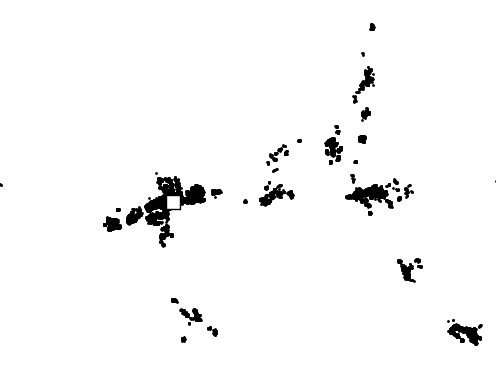}
         \caption[]{Valle-D-Aosta}
     \end{subfigure}
     \begin{subfigure}[b]{0.33\textwidth}
         \includegraphics[width=\textwidth]{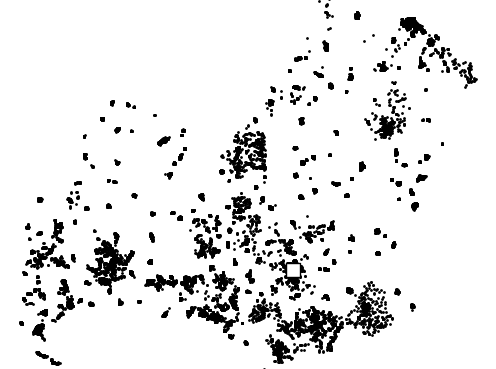}
         \caption[]{Molise}
     \end{subfigure}
	  \begin{subfigure}[b]{0.33\textwidth}
	\includegraphics[width=\textwidth]{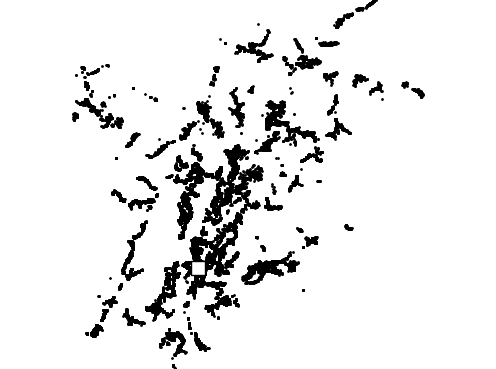}
	\caption[]{Trentino-Alto-Adige}
\end{subfigure}    

\bigskip

     \begin{subfigure}[b]{0.33\textwidth}
         \centering
         \includegraphics[width=\textwidth]{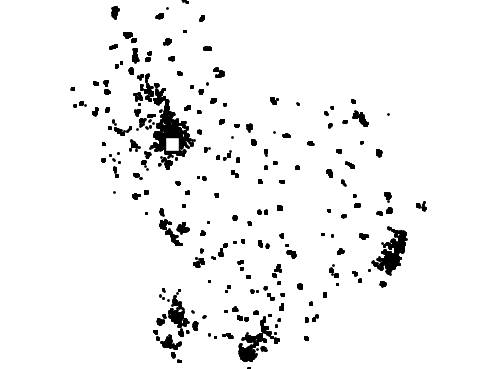}
         \caption[]{Basilicata}
     \end{subfigure}
     \begin{subfigure}[b]{0.33\textwidth}
	\includegraphics[width=\textwidth]{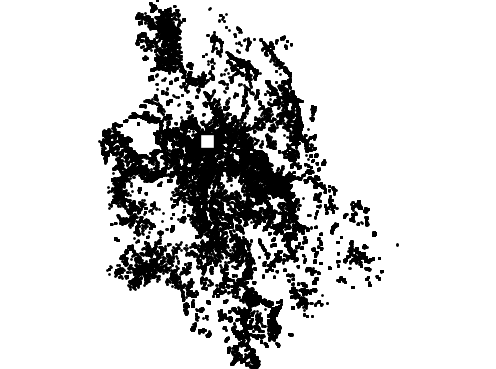}
	\caption[]{Umbria}
\end{subfigure}
\begin{subfigure}[b]{0.33\textwidth}
	\centering
	\includegraphics[width=\textwidth]{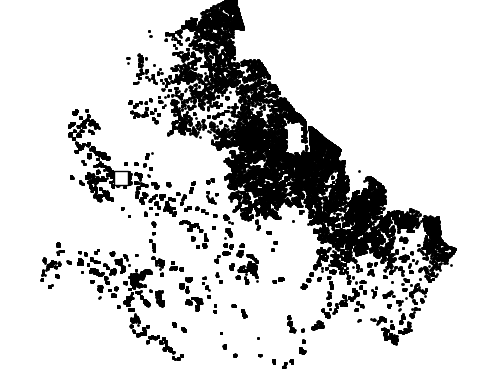}
	\caption[]{Abruzzo}
\end{subfigure}
     
\bigskip

     \begin{subfigure}[b]{0.33\textwidth}
         \includegraphics[width=\textwidth]{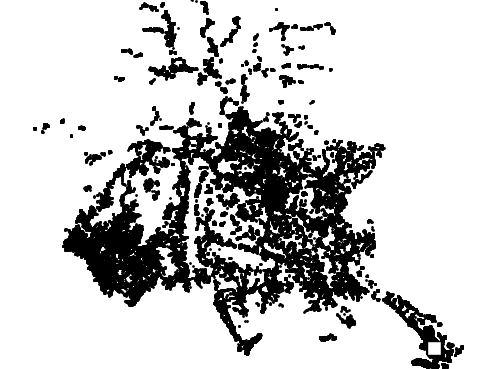}
         \caption[]{Friuli-Venezia-Giulia}
     \end{subfigure}
     \begin{subfigure}[b]{0.33\textwidth}
         \centering
         \includegraphics[width=\textwidth]{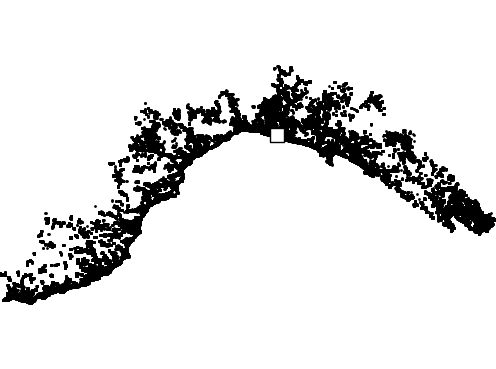}
         \caption[]{Liguria}
     \end{subfigure}
     \begin{subfigure}[b]{0.33\textwidth}
 	\centering
 	\includegraphics[width=\textwidth]{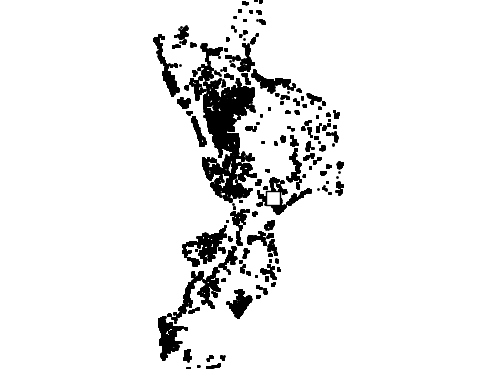}
 	\caption[]{Calabria}
 \end{subfigure}
 
 	\bigskip
 
 \begin{subfigure}[b]{0.33\textwidth}
 	\includegraphics[width=\textwidth]{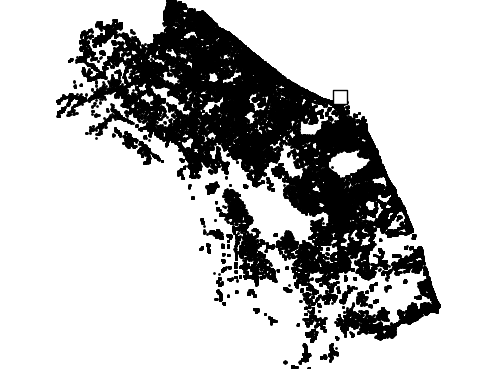}
 	\caption[]{Marche}
 \end{subfigure}
 \begin{subfigure}[b]{0.33\textwidth}
 	\centering
 	\includegraphics[width=\textwidth]{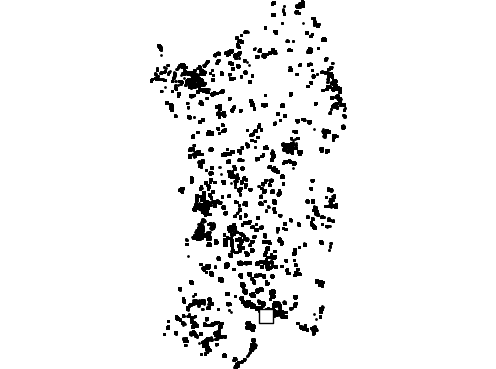}
 	\caption[]{Sardegna}
 \end{subfigure}
 \begin{subfigure}[b]{0.33\textwidth}
 	\centering
 	\includegraphics[width=\textwidth]{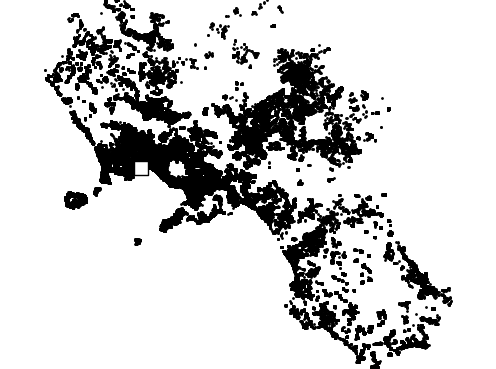}
 	\caption[]{Campania}
 \end{subfigure} 
 
\caption{$\mathbb{I}$ instances layout}
\label{fig:layout1}
\end{figure}

\begin{figure}[!t]
	\captionsetup[subfigure]{labelformat=empty}
	\footnotesize
	\centering

	\begin{subfigure}[b]{0.33\textwidth}
		\includegraphics[width=\textwidth]{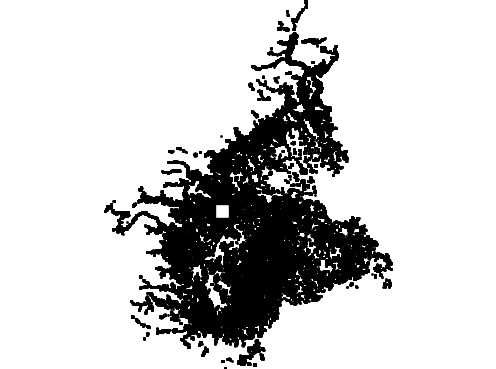}
		\caption[]{Piemonte}
	\end{subfigure}
	\begin{subfigure}[b]{0.33\textwidth}
		\includegraphics[width=\textwidth]{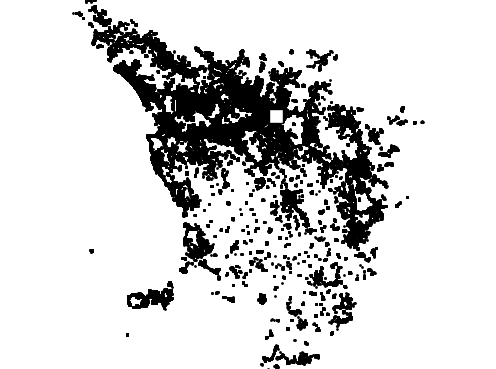}
		\caption[]{Toscana}
	\end{subfigure}
	\begin{subfigure}[b]{0.33\textwidth}
		\includegraphics[width=\textwidth]{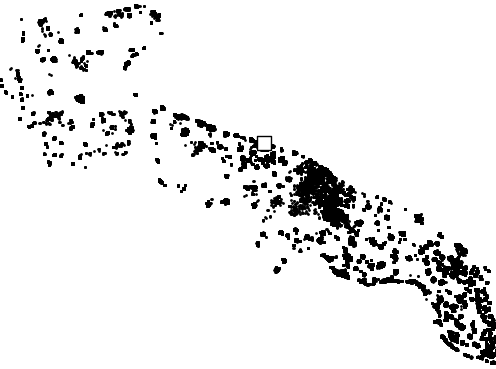}
		\caption[]{Puglia}
	\end{subfigure}    
	
	\bigskip

	\begin{subfigure}[b]{0.33\textwidth}
	\includegraphics[width=\textwidth]{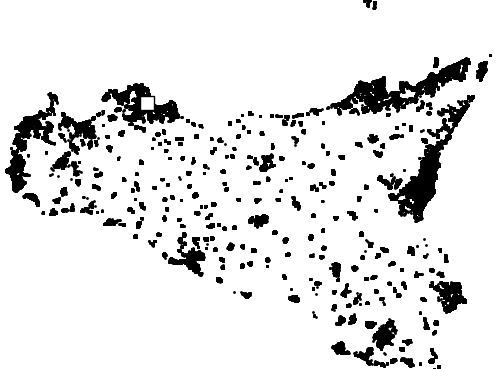}
	\caption[]{Sicilia}
\end{subfigure}
\begin{subfigure}[b]{0.33\textwidth}
	\includegraphics[width=\textwidth]{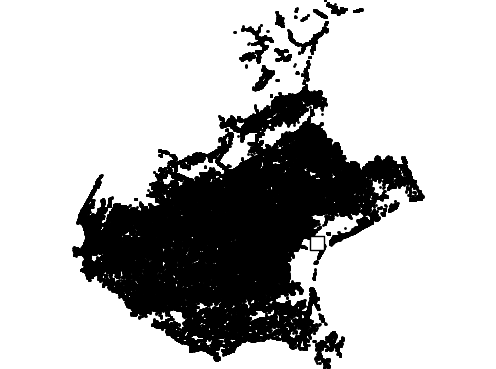}
	\caption[]{Veneto}
\end{subfigure}
\begin{subfigure}[b]{0.33\textwidth}
	\includegraphics[width=\textwidth]{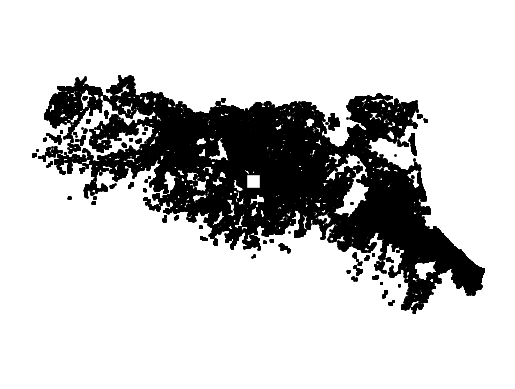}
	\caption[]{Emilia-Romagna}
\end{subfigure}    

\bigskip	

\begin{subfigure}[b]{0.33\textwidth}
	\includegraphics[width=\textwidth]{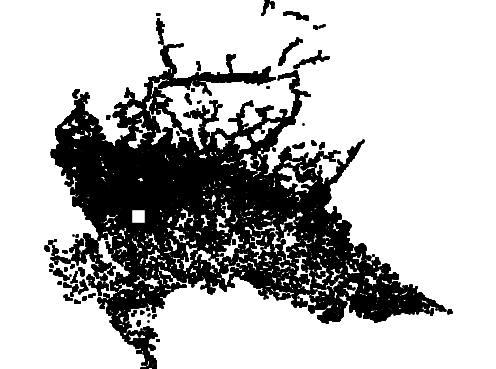}
	\caption[]{Lombardia}
\end{subfigure}    
\begin{subfigure}[b]{0.33\textwidth}
	\includegraphics[width=\textwidth]{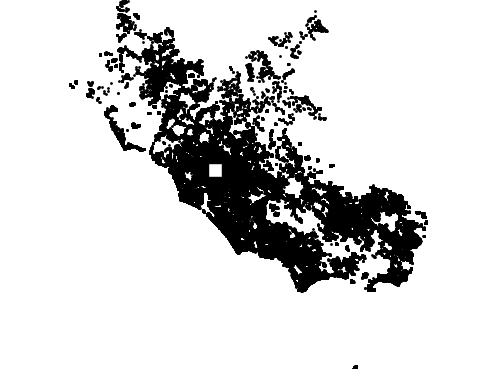}
	\caption[]{Lazio}
\end{subfigure}    

\caption{$\mathbb{I}$ instances layout}
\label{fig:layout2}

\end{figure}

Figures \ref{fig:layout1} and \ref{fig:layout2} show the layout of the new instances composing the $\mathbb{I}$ dataset. In particular, the black dots represent customers, while the depot is the empty square.

\clearpage

\section{Computational Details}
Sections \ref{appendix:x_instances}, \ref{appendix:b_instances}, and \ref{appendix:i_instances} provide additional computational details about the $\mathbb{X}$, $\mathbb{B}$, and $\mathbb{I}$ datasets, respectively.
For the $\mathbb{X}$ and $\mathbb{B}$ computations we also provide results of a statistical test to better evaluate whether there is some significant difference between the behavior of FILO2 and FILO.
More precisely, similarly to \citet{filo}, we performed a one-tailed Wilcoxon signed-rank test (\cite{wilcoxon}) where the null and alternative hypotheses $H_0$ and $H_1$, respectively, are defined as follows.
\begin{equation*}
    H_0: \Call{AverageCost}{FILO2} = \Call{AverageCost}{FILO}
\end{equation*}
\begin{equation*}
    H_1: \Call{AverageCost}{FILO2} > \Call{AverageCost}{FILO}    
\end{equation*}
A hypothesis is rejected when its associated $p$-value is lower than a significance level $\alpha$. 
Failing to reject $H_0$ means that the average results of FILO2 and FILO are not statistically different. 
On the other hand, when $H_0$ is rejected, the average results obtained by the algorithms are statistically different and the alternative hypothesis $H_1$ is tested to find whether the average results obtained by FILO2 are statistically greater than those of FILO. Rejecting $H_1$ implies that FILO2 performs better than FILO.

When performing multiple comparisons involving the same data, the probability of erroneously rejecting a null hypothesis increases. To control these errors, the significance level $\alpha$ is typically adjusted to lower values. Bonferroni correction (\cite{dunn}) is a simple method used to adjust $\alpha$ when performing multiple comparisons. In particular, given $n$ comparisons, the significance level is set to $\alpha / n$. 

\subsection{Computational Details for \texorpdfstring{$\mathbb{X}$}{X} Instances} \label
{appendix:x_instances}
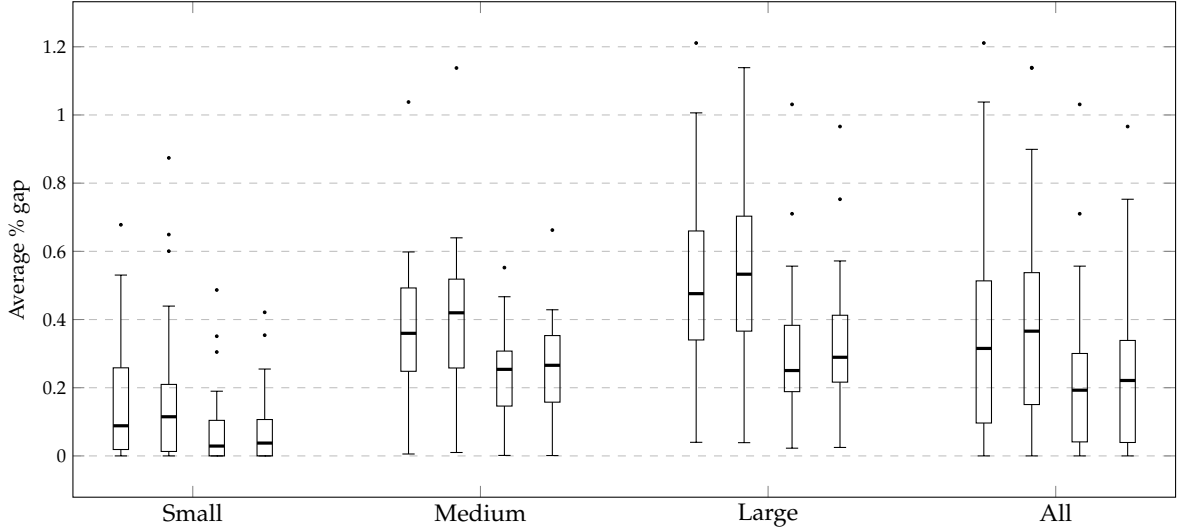
\begin{figure}
	\centering
	\scriptsize
\makeatletter
\pgfplotsset{
    boxplot/draw/average/.code={ 
        \draw [/pgfplots/boxplot/every average/.try]
        \pgfextra
        \pgftransformshift{%
            \pgfplotsboxplotpointabbox
            {\pgfplotsboxplotvalue{average}}
            {0.5}%
        }%
        \endpgfextra
        ;
    }
}
\makeatother
\begin{tikzpicture}
\begin{axis}
[
clip=false,
boxplot/draw direction=y,
boxplot/variable width,
boxplot/every median/.style={black,very thick,solid},
xmin=0.5,xmax=12,
width=1.00\textwidth,
height=0.33\textheight,
ylabel style={align=center}, 
y tick label style={align=right},
x tick label style={align=center},
xtick={1.75, 4.75, 7.75, 10.75},
ymajorgrids=true,
grid style=dashed,
xticklabels={{\small Small}, {\small Medium}, {\small Large}, {\small All}}, xticklabel style={rotate=0}, ylabel={\footnotesize Average \% gap},
scatter/classes={ a={mark=star}, b={mark=*}}]

\addplot[
mark=*, mark size=0.5pt,boxplot,
boxplot prepared={draw position=1, median=0.08841175892733341,upper quartile=0.2583770819744142,lower quartile=0.018772732605894214,upper whisker=0.5304039059032352,lower whisker=0.0,sample size=1}
] coordinates{ (1,0.67795246) };

\addplot[
mark=*, mark size=0.5pt,boxplot,
boxplot prepared={draw position=1.5, median=0.11486387132107562,upper quartile=0.20969879677156916,lower quartile=0.013071756838759864,upper whisker=0.43941757670306814,lower whisker=0.0,sample size=1}
] coordinates{ (1.5,0.60047894) (1.5,0.87404828) (1.5,0.64924612)};

\addplot[
mark=*, mark size=0.5pt,boxplot,
boxplot prepared={draw position=2, median=0.028893309063600277,upper quartile=0.1042732327139069,lower quartile=0.0,upper whisker=0.1897568165070041,lower whisker=0.0, sample size=1}
] coordinates{ (2, 0.30472191) (2, 0.35101404) (2, 0.48666631) };

\addplot[
mark=*, mark size=0.5pt,boxplot,
boxplot prepared={draw position=2.5,median=0.037658963983907875,upper quartile=0.10673201410198271,lower quartile=0.0,upper whisker=0.2548101924076963,lower whisker=0.0,sample size=1}
] coordinates{ (2.5, 0.35405784) (2.5, 0.42120513) };


\addplot[
mark=*, mark size=0.5pt,boxplot,
boxplot prepared={draw position=4, median=0.3595617206094358,upper quartile=0.49261969205594147,lower quartile=0.24816424161710016,upper whisker=0.5982443249932353,lower whisker=0.005615468062017968,sample size=1}
] coordinates{ (4, 1.03793999) };

\addplot[
mark=*, mark size=0.5pt,boxplot,
boxplot prepared={draw position=4.5, median=0.4197572967797989,upper quartile=0.5184471670059165,lower quartile=0.2578776503390728,upper whisker=0.6396538301377852,lower whisker=0.010019429569495142,sample size=1}
] coordinates{ (4.5, 1.13774885) };

\addplot[
mark=*, mark size=0.5pt,boxplot,
boxplot prepared={draw position=5, median=0.2538680071338787,upper quartile=0.3074479815692865,lower quartile=0.14619010838279742,upper whisker=0.4668674698795237,lower whisker=0.001403867015513778,sample size=1}
] coordinates{ (6,  0.55228469) };

\addplot[
mark=*, mark size=0.5pt,boxplot,
boxplot prepared={draw position=5.5, median=0.26575096188695346,upper quartile=0.35293463689254223,lower quartile=0.15764631342557694,upper whisker=0.42863762743280814,lower whisker=0.0010209941930992411,sample size=1}
] coordinates{ (6.5, 0.66237197) };


\addplot[
mark=*, mark size=0.5pt,boxplot,
boxplot prepared={draw position=7,median=0.47591369763919267,upper quartile=0.6597503574828602,lower quartile=0.34008466945066584,upper whisker=1.0061693114577779,lower whisker=0.039895111444096104, sample size=1}
] coordinates{ (7, 1.21101331) };

\addplot[
mark=*, mark size=0.5pt,boxplot,
boxplot prepared={draw position=7.5, median=0.5329374872717032,upper quartile=0.7031412605228039,lower quartile=0.3659763385500556,upper whisker=1.1387119700407213,lower whisker=0.038864955984266716,sample size=1}
] coordinates{ };

\addplot[
mark=*, mark size=0.5pt,boxplot,
boxplot prepared={draw position=8, median=0.25039858998451203,upper quartile=0.3829872497027471,lower quartile=0.18849391607462712,upper whisker=0.5565990602133846,lower whisker=0.02266342011612389,sample size=1}
] coordinates{ (8, 0.71020391) (8, 1.03101167) };

\addplot[
mark=*, mark size=0.5pt,boxplot,
boxplot prepared={draw position=8.5, median=0.2892843213391503,upper quartile=0.4125394681800303,lower quartile=0.2163733195204549,upper whisker=0.5718405524882917,lower whisker=0.02481738153212212,sample size=1}
] coordinates{ (8.5, 0.75279927) (8.5, 0.96609081) };


\addplot[
mark=*, mark size=0.5pt,boxplot,
boxplot prepared={draw position=10, median=0.3152217298847163,upper quartile=0.5133996928363967,lower quartile=0.09646720747594363,upper whisker=1.0379399884593166,lower whisker=0.0,sample size=1}
] coordinates{ (10, 1.21101331) };

\addplot[
mark=*, mark size=0.5pt,boxplot,
boxplot prepared={draw position=10.5, median=0.36588001513022594,upper quartile=0.5374862434114789,lower quartile=0.1506268482190573,upper whisker=0.8991447839354921,lower whisker=0.0,sample size=1}
] coordinates{ (10.5, 1.13774885) (10.5, 1.13871197) };

\addplot[
mark=*, mark size=0.5pt,boxplot,
boxplot prepared={draw position=11, median=0.19274829886950853,upper quartile=0.30043959282328886,lower quartile=0.04110905981869362,upper whisker=0.5565990602133846,lower whisker=0.0,sample size=1}
] coordinates{ (11, 0.71020391) (11, 1.03101167) };

\addplot[
mark=*, mark size=0.5pt,boxplot,
boxplot prepared={draw position=11.5, median=0.2212181348665398,upper quartile=0.3386658921633341,lower quartile=0.03936959608378905,upper whisker=0.7527992746135841,lower whisker=0.0,sample size=1}
] coordinates{ (11.5, 0.96609081) };

\end{axis}
\end{tikzpicture}
	\caption{Comparison of average gaps obtained by the FILO2 and FILO versions on the $\mathbb{X}$ dataset. For each group, boxplots, from left to right, are associated with: FILO2, FILO, FILO2 (long), and FILO (long).}
	\label{figure:x_box_plots}
\end{figure}

Detailed results about computations on the $\mathbb{X}$ dataset can be found in Tables \ref{table:x_small} -- \ref{table:x_large} and Figure \ref{figure:x_box_plots}.
\begin{sidewaystable}
    \centering
    \caption{Computations on small-sized $\mathbb{X}$ instances.\label{table:x_small}}
\footnotesize
\centering
\begin{tabular}{l r  rrrrrrrrrrrrrrrrrrrrr}
\toprule
& && \multicolumn{4}{c}{FILO} && \multicolumn{4}{c}{FILO2} && \multicolumn{4}{c}{FILO (long)} && \multicolumn{4}{c}{FILO2 (long)}\\
\cmidrule{4-7}
\cmidrule{9-12}
\cmidrule{14-17}
\cmidrule{19-22}

ID & BKS && Best & Avg & Worst & $t$  && Best & Avg & Worst & $t$ && Best & Avg & Worst & $t$ && Best & Avg & Worst & $t$ \\
\midrule
X-n101-k25&27591&&      0.00&      0.00&      0.00&        53&&      0.00&      0.00&      0.00&        49&&      0.00&      0.00&      0.00&       569&&      0.00&      0.00&      0.00&       534\\
X-n106-k14&26362&&      0.00&      0.06&      0.10&        94&&      0.00&      0.05&      0.11&        88&&      0.00&      0.02&      0.08&       922&&      0.00&      0.00&      0.03&       891\\
X-n110-k13&14971&&      0.00&      0.00&      0.00&        76&&      0.00&      0.00&      0.00&        78&&      0.00&      0.00&      0.00&       767&&      0.00&      0.00&      0.00&       782\\
X-n115-k10&12747&&      0.00&      0.00&      0.00&        76&&      0.00&      0.00&      0.00&        72&&      0.00&      0.00&      0.00&       784&&      0.00&      0.00&      0.00&       733\\
X-n120-k6&13332&&      0.00&      0.00&      0.00&        79&&      0.00&      0.00&      0.00&        83&&      0.00&      0.00&      0.00&       813&&      0.00&      0.00&      0.00&       849\\
X-n125-k30&55539&&      0.27&      0.60&      0.86&        58&&      0.08&      0.52&      0.86&        56&&      0.00&      0.10&      0.32&       562&&      0.00&      0.10&      0.29&       523\\
X-n129-k18&28940&&      0.03&      0.06&      0.19&        76&&      0.00&      0.07&      0.15&        68&&      0.00&      0.03&      0.05&       825&&      0.00&      0.02&      0.05&       764\\
X-n134-k13&10916&&      0.00&      0.15&      0.26&        72&&      0.00&      0.15&      0.22&        71&&      0.00&      0.06&      0.16&       776&&      0.00&      0.04&      0.16&       764\\
X-n139-k10&13590&&      0.00&      0.00&      0.00&        90&&      0.00&      0.00&      0.00&        95&&      0.00&      0.00&      0.00&       998&&      0.00&      0.00&      0.00&       998\\
X-n143-k7&15700&&      0.00&      0.15&      0.17&        68&&      0.00&      0.15&      0.17&        68&&      0.00&      0.13&      0.17&       787&&      0.00&      0.08&      0.17&       774\\
X-n148-k46&43448&&      0.00&      0.16&      0.38&        63&&      0.00&      0.16&      0.48&        54&&      0.00&      0.06&      0.30&       703&&      0.00&      0.09&      0.31&       632\\
X-n153-k22&21220&&      0.02&      0.21&      0.55&        79&&      0.02&      0.30&      0.69&        66&&      0.02&      0.04&      0.16&       668&&      0.02&      0.02&      0.02&       586\\
X-n157-k13&16876&&      0.00&      0.00&      0.00&       119&&      0.00&      0.00&      0.00&       112&&      0.00&      0.00&      0.00&      1311&&      0.00&      0.00&      0.00&      1334\\
X-n162-k11&14138&&      0.08&      0.18&      0.23&        84&&      0.00&      0.09&      0.22&        75&&      0.06&      0.10&      0.13&       861&&      0.00&      0.05&      0.08&       865\\
X-n167-k10&20557&&      0.00&      0.04&      0.13&        81&&      0.00&      0.10&      0.16&        85&&      0.00&      0.00&      0.00&       935&&      0.00&      0.00&      0.00&       844\\
X-n172-k51&45607&&      0.00&      0.01&      0.15&        59&&      0.00&      0.00&      0.00&        52&&      0.00&      0.00&      0.00&       614&&      0.00&      0.00&      0.00&       583\\
X-n176-k26&47812&&      0.03&      0.39&      0.91&        59&&      0.00&      0.50&      1.14&        54&&      0.00&      0.08&      0.28&       648&&      0.00&      0.13&      0.71&       654\\
X-n181-k23&25569&&      0.00&      0.01&      0.08&       101&&      0.00&      0.03&      0.09&       104&&      0.00&      0.00&      0.01&      1097&&      0.00&      0.00&      0.00&      1033\\
X-n186-k15&24145&&      0.02&      0.07&      0.16&        68&&      0.01&      0.05&      0.21&        64&&      0.01&      0.03&      0.12&       711&&      0.00&      0.05&      0.14&       704\\
X-n190-k8&16980&&      0.02&      0.08&      0.20&        78&&      0.02&      0.05&      0.11&        77&&      0.00&      0.02&      0.04&       788&&      0.00&      0.02&      0.04&       809\\
X-n195-k51&44225&&      0.06&      0.19&      0.43&        61&&      0.06&      0.22&      0.43&        52&&      0.00&      0.04&      0.19&       581&&      0.00&      0.07&      0.23&       527\\
X-n200-k36&58578&&      0.14&      0.87&      1.93&        66&&      0.45&      0.53&      0.79&        57&&      0.08&      0.35&      0.49&       750&&      0.02&      0.30&      0.49&       642\\
X-n204-k19&19565&&      0.03&      0.07&      0.20&        63&&      0.00&      0.09&      0.68&        62&&      0.00&      0.00&      0.03&       662&&      0.00&      0.00&      0.00&       693\\
X-n209-k16&30656&&      0.07&      0.16&      0.46&        59&&      0.01&      0.09&      0.14&        58&&      0.00&      0.05&      0.09&       612&&      0.00&      0.03&      0.09&       599\\
X-n214-k11&10856&&      0.13&      0.35&      0.60&        58&&      0.05&      0.26&      0.49&        60&&      0.04&      0.18&      0.38&       633&&      0.10&      0.19&      0.30&       614\\
X-n219-k73&117595&&      0.00&      0.00&      0.01&       227&&      0.00&      0.00&      0.00&       231&&      0.00&      0.00&      0.00&      2489&&      0.00&      0.00&      0.00&      2493\\
X-n223-k34&40437&&      0.08&      0.21&      0.30&        55&&      0.09&      0.31&      0.61&        58&&      0.00&      0.17&      0.25&       586&&      0.16&      0.18&      0.25&       549\\
X-n228-k23&25742&&      0.00&      0.20&      0.41&        59&&      0.17&      0.22&      0.38&        60&&      0.00&      0.17&      0.21&       618&&      0.00&      0.14&      0.20&       656\\
X-n233-k16&19230&&      0.16&      0.44&      0.56&        75&&      0.00&      0.44&      0.57&        65&&      0.06&      0.25&      0.53&       748&&      0.00&      0.35&      0.53&       746\\
X-n237-k14&27042&&      0.00&      0.01&      0.03&        90&&      0.00&      0.06&      0.36&        96&&      0.00&      0.04&      0.16&      1043&&      0.00&      0.01&      0.09&      1116\\
X-n242-k48&82751&&      0.09&      0.30&      0.70&        77&&      0.03&      0.27&      0.41&        67&&      0.09&      0.19&      0.24&       851&&      0.04&      0.15&      0.33&       778\\
X-n247-k50&37274&&      0.07&      0.65&      0.87&        68&&      0.02&      0.68&      1.03&        66&&      0.01&      0.42&      0.75&       757&&      0.03&      0.49&      0.73&       769\\
\midrule
Mean&&&      0.04&      0.18&      0.34&        78&&      0.03&      0.17&      0.33&        75&&      0.01&      0.08&      0.16&       827&&      0.01&      0.08&      0.16&       807\\
\bottomrule
\end{tabular}
\end{sidewaystable}
\begin{sidewaystable}
    \centering
    \caption{Computations on medium-sized $\mathbb{X}$ instances.\label{table:x_medium}}
\footnotesize
\centering
\begin{tabular}{l r  rrrrrrrrrrrrrrrrrrrrr}
\toprule
& && \multicolumn{4}{c}{FILO} && \multicolumn{4}{c}{FILO2} && \multicolumn{4}{c}{FILO (long)} && \multicolumn{4}{c}{FILO2 (long)}\\
\cmidrule{4-7}
\cmidrule{9-12}
\cmidrule{14-17}
\cmidrule{19-22}

ID & BKS && Best & Avg & Worst & $t$  && Best & Avg & Worst & $t$ && Best & Avg & Worst & $t$ && Best & Avg & Worst & $t$ \\
\midrule
X-n251-k28&38684&&      0.04&      0.34&      0.44&        69&&      0.21&      0.37&      0.51&        74&&      0.16&      0.31&      0.40&       812&&      0.17&      0.26&      0.37&       750\\
X-n256-k16&18839&&      0.22&      0.22&      0.22&        92&&      0.22&      0.22&      0.22&        89&&      0.22&      0.22&      0.22&       994&&      0.22&      0.22&      0.22&       970\\
X-n261-k13&26558&&      0.28&      0.48&      0.71&        64&&      0.23&      0.36&      0.72&        64&&      0.01&      0.36&      0.50&       745&&      0.01&      0.27&      0.50&       692\\
X-n266-k58&75478&&      0.21&      0.52&      0.72&        91&&      0.17&      0.36&      0.52&        76&&      0.18&      0.42&      0.54&       960&&      0.19&      0.37&      0.61&       941\\
X-n270-k35&35291&&      0.09&      0.24&      0.35&        60&&      0.05&      0.22&      0.35&        61&&      0.07&      0.16&      0.28&       657&&      0.05&      0.14&      0.27&       640\\
X-n275-k28&21245&&      0.00&      0.06&      0.31&        81&&      0.00&      0.07&      0.29&        93&&      0.00&      0.00&      0.04&       884&&      0.00&      0.01&      0.12&       994\\
X-n280-k17&33503&&      0.29&      0.48&      0.67&        63&&      0.12&      0.28&      0.45&        60&&      0.16&      0.39&      0.53&       661&&      0.06&      0.31&      0.50&       743\\
X-n284-k15&20215&&      0.17&      0.46&      0.82&        69&&      0.34&      0.51&      0.87&        69&&      0.09&      0.22&      0.31&       684&&      0.10&      0.27&      0.46&       669\\
X-n289-k60&95151&&      0.46&      0.59&      0.75&        86&&      0.51&      0.59&      0.86&        74&&      0.28&      0.38&      0.48&       890&&      0.26&      0.39&      0.50&       847\\
X-n294-k50&47161&&      0.22&      0.35&      0.45&        51&&      0.14&      0.30&      0.44&        50&&      0.17&      0.27&      0.35&       542&&      0.15&      0.20&      0.34&       478\\
X-n298-k31&34231&&      0.13&      0.27&      0.39&        53&&      0.01&      0.26&      0.59&        52&&      0.18&      0.23&      0.27&       553&&      0.00&      0.16&      0.26&       555\\
X-n303-k21&21736&&      0.31&      0.48&      1.01&        53&&      0.27&      0.41&      0.68&        55&&      0.10&      0.34&      0.44&       514&&      0.12&      0.32&      0.47&       539\\
X-n308-k13&25859&&      0.02&      0.35&      1.63&        67&&      0.07&      0.60&      1.43&        68&&      0.03&      0.33&      0.72&       799&&      0.04&      0.27&      0.78&       735\\
X-n313-k71&94043&&      0.26&      0.58&      0.80&        66&&      0.38&      0.53&      0.75&        61&&      0.19&      0.29&      0.52&       683&&      0.22&      0.30&      0.49&       663\\
X-n317-k53&78355&&      0.00&      0.01&      0.04&       135&&      0.00&      0.01&      0.03&       142&&      0.00&      0.00&      0.01&      1442&&      0.00&      0.00&      0.01&      1464\\
X-n322-k28&29834&&      0.25&      0.53&      0.92&        54&&      0.16&      0.34&      0.69&        56&&      0.05&      0.36&      0.62&       608&&      0.11&      0.35&      0.49&       598\\
X-n327-k20&27532&&      0.20&      0.44&      0.74&        71&&      0.13&      0.39&      0.74&        74&&      0.09&      0.24&      0.37&       809&&      0.00&      0.26&      0.32&       800\\
X-n331-k15&31102&&      0.00&      0.05&      0.31&        85&&      0.00&      0.01&      0.01&        86&&      0.00&      0.01&      0.01&       919&&      0.00&      0.00&      0.01&       962\\
X-n336-k84&139111&&      0.49&      0.63&      0.90&        67&&      0.49&      0.57&      0.72&        57&&      0.20&      0.35&      0.48&       643&&      0.18&      0.35&      0.51&       595\\
X-n344-k43&42050&&      0.34&      0.56&      0.70&        60&&      0.23&      0.49&      0.72&        61&&      0.20&      0.36&      0.52&       555&&      0.16&      0.31&      0.43&       590\\
X-n351-k40&25896&&      0.37&      0.54&      0.76&        59&&      0.44&      0.60&      0.73&        59&&      0.27&      0.43&      0.66&       567&&      0.27&      0.47&      0.61&       610\\
X-n359-k29&51505&&      0.30&      0.47&      0.69&        64&&      0.17&      0.44&      0.84&        62&&      0.09&      0.24&      0.38&       726&&      0.04&      0.15&      0.32&       683\\
X-n367-k17&22814&&      0.00&      0.19&      0.44&        68&&      0.00&      0.12&      0.50&        70&&      0.00&      0.01&      0.06&       710&&      0.00&      0.02&      0.05&       738\\
X-n376-k94&147713&&      0.00&      0.01&      0.03&       180&&      0.00&      0.01&      0.02&       173&&      0.00&      0.00&      0.02&      1894&&      0.00&      0.00&      0.01&      1792\\
X-n384-k52&65928&&      0.24&      0.43&      0.72&        74&&      0.27&      0.42&      0.53&        72&&      0.18&      0.32&      0.38&       835&&      0.15&      0.25&      0.42&       783\\
X-n393-k38&38260&&      0.08&      0.28&      0.58&        59&&      0.10&      0.24&      0.60&        65&&      0.08&      0.12&      0.27&       656&&      0.02&      0.09&      0.12&       653\\
X-n401-k29&66154&&      0.11&      0.26&      0.39&        85&&      0.18&      0.25&      0.32&        86&&      0.11&      0.15&      0.19&       858&&      0.09&      0.13&      0.15&       865\\
X-n411-k19&19712&&      0.27&      0.50&      0.98&        74&&      0.26&      0.41&      0.56&        77&&      0.25&      0.35&      0.60&       795&&      0.31&      0.33&      0.36&       840\\
X-n420-k130&107798&&      0.13&      0.22&      0.31&        53&&      0.12&      0.27&      0.36&        53&&      0.09&      0.14&      0.27&       511&&      0.03&      0.15&      0.25&       540\\
X-n429-k61&65449&&      0.28&      0.40&      0.53&        62&&      0.23&      0.41&      0.71&        61&&      0.12&      0.23&      0.36&       674&&      0.15&      0.20&      0.29&       648\\
X-n439-k37&36391&&      0.01&      0.10&      0.24&        71&&      0.01&      0.07&      0.24&        80&&      0.01&      0.01&      0.01&       767&&      0.01&      0.01&      0.04&       821\\
X-n449-k29&55233&&      0.41&      0.64&      0.89&        61&&      0.30&      0.60&      0.92&        64&&      0.24&      0.38&      0.59&       652&&      0.17&      0.29&      0.43&       634\\
X-n459-k26&24139&&      0.22&      0.41&      0.83&        67&&      0.12&      0.26&      0.46&        69&&      0.12&      0.26&      0.42&       735&&      0.14&      0.25&      0.43&       732\\
X-n469-k138&221824&&      0.93&      1.14&      1.31&        66&&      0.76&      1.04&      1.23&        73&&      0.49&      0.66&      0.82&       669&&      0.42&      0.55&      0.68&       681\\
X-n480-k70&89449&&      0.24&      0.39&      0.49&        71&&      0.24&      0.32&      0.43&        72&&      0.05&      0.22&      0.31&       746&&      0.02&      0.18&      0.45&       799\\
X-n491-k59&66483&&      0.28&      0.53&      0.69&        62&&      0.39&      0.53&      0.74&        62&&      0.16&      0.35&      0.49&       610&&      0.25&      0.31&      0.39&       636\\
\midrule
Mean&&&      0.22&      0.39&      0.63&        73&&      0.20&      0.36&      0.58&        73&&      0.13&      0.25&      0.37&       771&&      0.11&      0.23&      0.35&       769\\
\bottomrule
\end{tabular}
\end{sidewaystable}
\begin{sidewaystable}
    \centering
    \caption{Computations on large-sized $\mathbb{X}$ instances.\label{table:x_large}}
\footnotesize
\centering
\begin{tabular}{l r  rrrrrrrrrrrrrrrrrrrrr}
\toprule
& && \multicolumn{4}{c}{FILO} && \multicolumn{4}{c}{FILO2} && \multicolumn{4}{c}{FILO (long)} && \multicolumn{4}{c}{FILO2 (long)}\\
\cmidrule{4-7}
\cmidrule{9-12}
\cmidrule{14-17}
\cmidrule{19-22}

ID & BKS && Best & Avg & Worst & $t$  && Best & Avg & Worst & $t$ && Best & Avg & Worst & $t$ && Best & Avg & Worst & $t$ \\
\midrule
X-n502-k39&69226&&      0.00&      0.05&      0.09&       105&&      0.01&      0.06&      0.10&       120&&      0.00&      0.03&      0.07&      1089&&      0.01&      0.04&      0.06&      1222\\
X-n513-k21&24201&&      0.15&      0.37&      0.69&        73&&      0.00&      0.26&      0.45&        82&&      0.00&      0.13&      0.40&       762&&      0.00&      0.11&      0.32&       843\\
X-n524-k153&154593&&      0.27&      0.46&      0.59&        62&&      0.19&      0.44&      0.97&        67&&      0.04&      0.23&      0.39&       618&&      0.01&      0.20&      0.48&       670\\
X-n536-k96&94846&&      0.69&      0.87&      0.95&        70&&      0.76&      0.85&      0.92&        70&&      0.68&      0.75&      0.80&       708&&      0.56&      0.71&      0.81&       702\\
X-n548-k50&86700&&      0.06&      0.12&      0.19&        80&&      0.00&      0.09&      0.22&        92&&      0.00&      0.04&      0.12&       828&&      0.00&      0.06&      0.11&       956\\
X-n561-k42&42717&&      0.26&      0.45&      0.63&        57&&      0.18&      0.39&      0.62&        59&&      0.16&      0.31&      0.52&       578&&      0.10&      0.22&      0.31&       615\\
X-n573-k30&50673&&      0.24&      0.37&      0.51&        79&&      0.24&      0.35&      0.60&        88&&      0.17&      0.29&      0.48&       809&&      0.14&      0.23&      0.41&       923\\
X-n586-k159&190316&&      0.54&      0.64&      0.89&        74&&      0.38&      0.57&      0.77&        77&&      0.23&      0.38&      0.55&       790&&      0.21&      0.38&      0.47&       797\\
X-n599-k92&108451&&      0.37&      0.54&      0.69&        72&&      0.27&      0.37&      0.54&        78&&      0.24&      0.29&      0.40&       804&&      0.19&      0.27&      0.40&       831\\
X-n613-k62&59535&&      0.36&      0.63&      0.82&        51&&      0.47&      0.68&      0.84&        54&&      0.19&      0.43&      0.66&       500&&      0.08&      0.32&      0.52&       540\\
X-n627-k43&62164&&      0.22&      0.36&      0.52&        76&&      0.31&      0.37&      0.46&        79&&      0.09&      0.23&      0.33&       796&&      0.09&      0.21&      0.26&       791\\
X-n641-k35&63682&&      0.16&      0.42&      0.61&        79&&      0.24&      0.31&      0.44&        83&&      0.20&      0.27&      0.34&       829&&      0.12&      0.23&      0.43&       812\\
X-n655-k131&106780&&      0.02&      0.04&      0.06&       141&&      0.01&      0.04&      0.06&       162&&      0.00&      0.02&      0.04&      1479&&      0.00&      0.02&      0.04&      1633\\
X-n670-k130&146332&&      0.65&      1.14&      1.69&        64&&      0.53&      1.21&      2.21&        69&&      0.65&      0.97&      1.35&       663&&      0.62&      1.03&      1.83&       724\\
X-n685-k75&68205&&      0.47&      0.66&      1.05&        63&&      0.36&      0.62&      0.73&        67&&      0.32&      0.46&      0.58&       593&&      0.23&      0.44&      0.60&       685\\
X-n701-k44&81923&&      0.49&      0.54&      0.66&        67&&      0.23&      0.43&      0.57&        70&&      0.15&      0.28&      0.44&       672&&      0.12&      0.22&      0.32&       710\\
X-n716-k35&43373&&      0.60&      0.72&      0.89&        70&&      0.47&      0.62&      0.73&        75&&      0.25&      0.36&      0.54&       702&&      0.16&      0.28&      0.44&       742\\
X-n733-k159&136187&&      0.26&      0.40&      0.60&        60&&      0.23&      0.36&      0.48&        65&&      0.12&      0.22&      0.34&       572&&      0.12&      0.20&      0.29&       661\\
X-n749-k98&77269&&      0.71&      0.80&      0.94&        65&&      0.62&      0.75&      0.89&        66&&      0.46&      0.52&      0.60&       608&&      0.30&      0.44&      0.56&       647\\
X-n766-k71&114417&&      0.43&      0.70&      0.96&        70&&      0.37&      0.58&      0.68&        74&&      0.26&      0.46&      0.62&       699&&      0.19&      0.45&      0.67&       746\\
X-n783-k48&72386&&      0.40&      0.68&      0.81&        76&&      0.23&      0.48&      0.87&        85&&      0.06&      0.29&      0.45&       794&&      0.18&      0.29&      0.40&       847\\
X-n801-k40&73305&&      0.10&      0.22&      0.34&        74&&      0.13&      0.25&      0.44&        89&&      0.09&      0.16&      0.27&       782&&      0.06&      0.12&      0.19&       893\\
X-n819-k171&158121&&      0.70&      0.85&      1.02&        64&&      0.62&      0.76&      0.95&        66&&      0.47&      0.57&      0.70&       650&&      0.48&      0.56&      0.61&       690\\
X-n837-k142&193737&&      0.46&      0.52&      0.63&        76&&      0.40&      0.48&      0.56&        81&&      0.19&      0.28&      0.40&       823&&      0.22&      0.29&      0.36&       861\\
X-n856-k95&88965&&      0.12&      0.18&      0.24&        80&&      0.09&      0.16&      0.25&       100&&      0.06&      0.10&      0.15&       801&&      0.02&      0.07&      0.13&      1001\\
X-n876-k59&99299&&      0.35&      0.45&      0.57&        77&&      0.38&      0.47&      0.59&        78&&      0.15&      0.26&      0.34&       744&&      0.14&      0.23&      0.35&       781\\
X-n895-k37&53860&&      0.51&      0.76&      1.03&        74&&      0.45&      0.61&      0.80&        83&&      0.18&      0.33&      0.47&       760&&      0.19&      0.33&      0.49&       852\\
X-n916-k207&329179&&      0.56&      0.67&      0.88&        86&&      0.54&      0.66&      0.81&        85&&      0.28&      0.41&      0.56&       833&&      0.29&      0.39&      0.52&       853\\
X-n936-k151&132715&&      0.69&      0.90&      1.05&        59&&      0.70&      0.93&      1.34&        71&&      0.38&      0.54&      0.78&       581&&      0.34&      0.53&      0.75&       690\\
X-n957-k87&85465&&      0.12&      0.17&      0.25&        81&&      0.05&      0.11&      0.21&       100&&      0.04&      0.11&      0.18&       835&&      0.06&      0.09&      0.12&      1021\\
X-n979-k58&118976&&      0.29&      0.44&      1.15&       102&&      0.80&      1.01&      1.12&       104&&      0.12&      0.20&      0.31&       999&&      0.13&      0.17&      0.20&      1064\\
X-n1001-k43&72355&&      0.53&      0.76&      1.07&        71&&      0.50&      0.66&      0.83&        78&&      0.20&      0.30&      0.46&       717&&      0.13&      0.27&      0.45&       777\\
\midrule
Mean&&&      0.37&      0.53&      0.72&        75&&      0.34&      0.50&      0.69&        82&&      0.20&      0.32&      0.46&       763&&      0.17&      0.29&      0.43&       831\\
\bottomrule
\end{tabular}
\end{sidewaystable}

\clearpage
Table \ref{table:p_values_x} shows the $p$-values of the Wilcoxon test on $\mathbb{X}$ computations. We considered a total number of $n=8$ hypotheses corresponding to the partitioning of instances (Small, Medium, Large, and All) and to the two hypotheses ($H_0$ and $H_1$). Thus, by assuming an initial significance level $\alpha_0 = 0.025$, the adjusted value becomes $\alpha=\alpha_0 / 8 = 0.003125$.
In particular we have that
\begin{itemize}
	\item FILO2 and FILO behave similarly on the small instances.
	\item FILO2 performs better than FILO on medium and large instances, and on the overall dataset.
	\item FILO2 (long) and FILO (long) behave similarly on the small instances.
	\item FILO2 (long) performs better than FILO (long) on medium and large instances, and on the overall dataset.
\end{itemize}

\begin{table}
	\caption{Computations on the $\mathbb{X}$ dataset: $p$-values for FILO2 vs FILO on the left and FILO2 (long) vs FILO (long) on the right.
	\\
	\footnotesize
	$p$-values in bold are associated with rejected hypothesis when $\alpha = 0.003125$.\\
The last row of each group contains a $p$-value interpretation when $\alpha = 0.003125$. In particular, FILO2 is not statistically different from FILO when $H_0$ cannot be rejected (Similar), FILO2 is statistically better when both $H_0$ and $H_1$ are rejected (Better), and, finally, FILO2 is statistically worse when $H_0$ is rejected and $H_1$ is not rejected (Worse).}
	\label{table:p_values_x}
	\centering
	\footnotesize
	\begin{tabular}{rrrrr}
	\toprule
		& \multicolumn{4}{c}{FILO2 vs FILO}\\
	\midrule
	& Small & Medium & Large & All \\
	\midrule
	$H_0$ & 0.809330 & \textbf{0.000786} & \textbf{0.002750} & \textbf{0.000030}\\
	$H_1$ & 0.404665 & \textbf{0.000393} & \textbf{0.001375} & \textbf{0.000015}\\
		\midrule
	& Similar  & Better & Better & Better\\
	\bottomrule
	\end{tabular}
	\qquad
	\begin{tabular}{rrrrr}
	\toprule
		& \multicolumn{4}{c}{FILO2 (long) vs FILO (long)}\\
	\midrule
	& Small & Medium & Large & All \\
	\midrule
	$H_0$ & 0.626246 & \textbf{0.000600} & \textbf{0.000178} & \textbf{0.000001}\\
	$H_1$ & 0.313123 & \textbf{0.000300} & \textbf{0.000089} & \textbf{0.000001}\\
		\midrule
	& Similar  & Better & Better & Better\\
	\bottomrule
	\end{tabular}
\end{table}

\clearpage
\subsection{Computational Details for \texorpdfstring{$\mathbb{B}$}{B} Instances} \label{appendix:b_instances}
Detailed results about computations on the $\mathbb{B}$ dataset can be found in Table \ref{table:b-details} and Figure \ref{figure:b_box_plots}.
\begin{figure}
	\centering
	\scriptsize
\makeatletter
\pgfplotsset{
    boxplot/draw/average/.code={ 
        \draw [/pgfplots/boxplot/every average/.try]
        \pgfextra
        \pgftransformshift{%
            \pgfplotsboxplotpointabbox
            {\pgfplotsboxplotvalue{average}}
            {0.5}%
        }%
        \endpgfextra
        ;
    }
}
\makeatother
\begin{tikzpicture}
\begin{axis}
[
clip=false,
boxplot/draw direction=y,
boxplot/variable width,
boxplot/every median/.style={black,very thick,solid},
xmin=0.5,xmax=3,
width=1.00\textwidth,
height=0.33\textheight,
ylabel style={align=center}, 
y tick label style={align=right},
x tick label style={align=center},
xtick={1, 1.5, 2, 2.5},
ymajorgrids=true,
grid style=dashed,
xticklabels={{\small FILO2}, {\small FILO}, {\small FILO2 (long)}, {\small FILO (long)}}, xticklabel style={rotate=0}, ylabel={\footnotesize Average \% gap},
scatter/classes={ a={mark=star}, b={mark=*}}]

\addplot[
mark=*, mark size=0.5pt,boxplot,
boxplot prepared={draw position=1, median=0.9354908739715571,upper quartile=1.374134705852478,lower quartile=0.675998179087498,upper whisker=2.125611663450189,lower whisker=0.4190865344727514,sample size=1}
] coordinates{};

\addplot[
mark=*, mark size=0.5pt,boxplot,
boxplot prepared={draw position=1.5, median=1.0535004728399668,upper quartile=1.4080556550807606,lower quartile=0.7579035918936734,upper whisker=2.1054827911060703,lower whisker=0.5307807184933242,sample size=1}
] coordinates{};

\addplot[
mark=*, mark size=0.5pt,boxplot,
boxplot prepared={draw position=2, median=0.3521703908030046,upper quartile=0.407360010515657,lower quartile=0.2815819512614564,upper whisker=0.5490605850973058,lower whisker=0.20332075002074473,sample size=1}
] coordinates{ (2, 0.61638755) };

\addplot[
mark=*, mark size=0.5pt,boxplot,
boxplot prepared={draw position=2.5, median=0.4275108698005593,upper quartile=0.4785747842624955,lower quartile=0.31167308848680597,upper whisker=0.6786057274564776,lower whisker=0.26440512735418265,sample size=1}
] coordinates{ };


\end{axis}
\end{tikzpicture}
	\caption{Comparison of average gaps obtained by algorithms on the $\mathbb{B}$ dataset.}
	\label{figure:b_box_plots}
\end{figure}
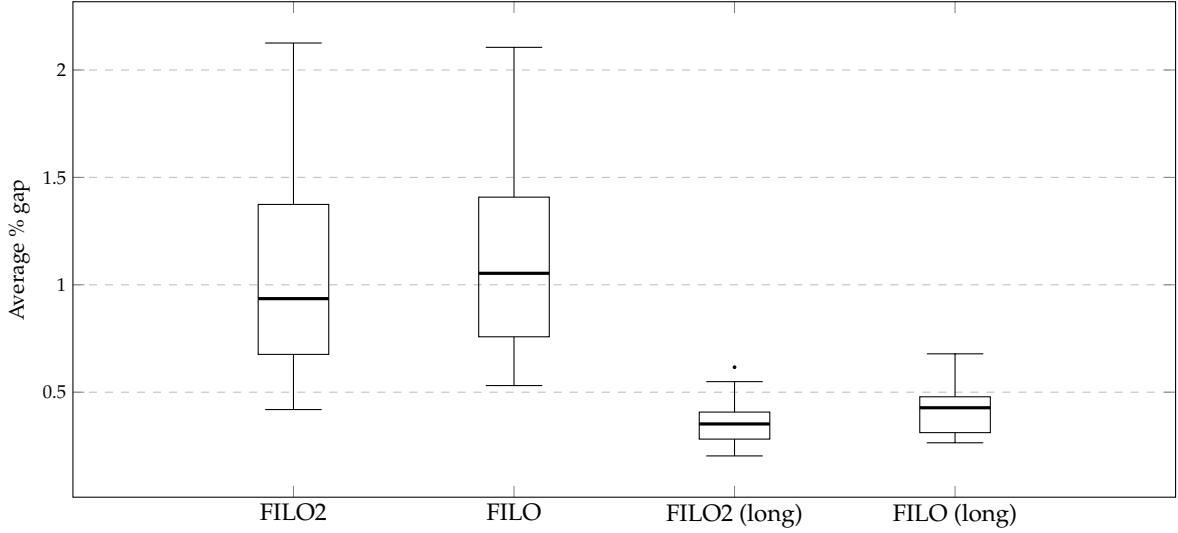
\begin{table}
    \centering
    \caption{Computations on $\mathbb{B}$ instances.}
	\label{table:b-details}
\footnotesize
\centering
\setlength\tabcolsep{4pt}
\begin{tabular}{l r  rrrrrrrrrrrrrrrrrrrrr}
\toprule
& && \multicolumn{4}{c}{FILO} && \multicolumn{4}{c}{FILO2} && \multicolumn{4}{c}{FILO (long)} && \multicolumn
{4}{c}{FILO2 (long)}\\
\cmidrule{4-7}
\cmidrule{9-12}
\cmidrule{14-17}
\cmidrule{19-22} 
ID & BKS && Best & Avg & Worst & t &&  Best & Avg & Worst & t &&  Best & Avg & Worst & t &&  Best & Avg & Worst & t \\
\midrule
L1 & 192848 &&       0.47 &       0.53 &       0.65 &         95 &&       0.35 &       0.42 &       0.50 &         98 &&       0.19 &       0.26 &       0.31 &        958 &&       0.16 &       0.20 &       0.24 &        992\\
L2 & 111391 &&       0.81 &       1.03 &       1.20 &        142 &&       0.57 &       0.85 &       1.34 &        130 &&       0.40 &       0.49 &       0.57 &       1515 &&       0.17 &       0.32 &       0.46 &       1479\\
A1 & 477277 &&       0.54 &       0.61 &       0.66 &        124 &&       0.49 &       0.52 &       0.56 &        105 &&       0.25 &       0.29 &       0.33 &       1247 &&       0.22 &       0.25 &       0.29 &       1073\\
A2 & 291350 &&       1.02 &       1.10 &       1.22 &        139 &&       0.96 &       1.08 &       1.27 &        114 &&       0.33 &       0.44 &       0.52 &       1505 &&       0.28 &       0.40 &       0.50 &       1226\\
G1 & 469531 &&       0.70 &       0.74 &       0.76 &        161 &&       0.60 &       0.66 &       0.72 &        114 &&       0.27 &       0.30 &       0.34 &       1602 &&       0.25 &       0.27 &       0.29 &       1180\\
G2 & 257748 &&       1.42 &       1.51 &       1.60 &        198 &&       1.30 &       1.47 &       1.61 &        127 &&       0.39 &       0.44 &       0.53 &       2284 &&       0.33 &       0.41 &       0.49 &       1505\\
B1 & 501719 &&       1.02 &       1.08 &       1.13 &        197 &&       0.94 &       1.02 &       1.10 &        113 &&       0.38 &       0.42 &       0.46 &       2006 &&       0.35 &       0.39 &       0.42 &       1206\\
B2 & 345481 &&       1.90 &       2.01 &       2.15 &        228 &&       1.85 &       1.91 &       2.01 &        122 &&       0.52 &       0.58 &       0.66 &       2564 &&       0.46 &       0.55 &       0.62 &       1430\\
F1 & 7240124 &&       0.77 &       0.81 &       0.85 &        303 &&       0.69 &       0.73 &       0.78 &        136 &&       0.32 &       0.35 &       0.38 &       3276 &&       0.30 &       0.31 &       0.33 &       1575\\
F2 & 4373320 &&       1.99 &       2.11 &       2.23 &        482 &&       1.94 &       2.13 &       2.21 &        153 &&       0.60 &       0.68 &       0.83 &       6194 &&       0.56 &       0.62 &       0.72 &       2040\\
\midrule
Mean &  &&       1.06 &       1.15 &       1.24 &        207 &&       0.97 &       1.08 &       1.21 &        121 &&       0.36 &       0.42 &       0.49 &       2315 &&       0.31 &       0.37 &       0.44 &       1371\\
\bottomrule
\end{tabular}
\end{table}

\clearpage
Table \ref{table:p_values_b} shows the $p$-values of the Wilcoxon test on $\mathbb{B}$ computations. We considered a total number of $n=2$ hypotheses corresponding to $H_0$ and $H_1$. Thus, by assuming an initial significance level $\alpha_0 = 0.025$, the adjusted value becomes $\alpha=\alpha_0 / 2 = 0.0125$.
In particular we have that
\begin{itemize}
	\item FILO2 and FILO behave similarly.
	\item FILO2 (long) performs better than FILO (long).
\end{itemize}
\begin{table}
	\caption{Computations on the $\mathbb{B}$ dataset: $p$-values for FILO2 vs FILO on the left and FILO2 (long) vs FILO (long) on the right.
	\\
	\footnotesize
	$p$-values in bold are associated with rejected hypothesis when $\alpha = 0.0125$.\\
The last row of each group contains a $p$-value interpretation when $\alpha = 0.0125$. In particular, FILO2 is not statistically different from FILO when $H_0$ cannot be rejected (Similar), FILO2 is statistically better when both $H_0$ and $H_1$ are rejected (Better), and, finally, FILO2 is statistically worse when $H_0$ is rejected and $H_1$ is not rejected (Worse).}
	\label{table:p_values_b}
	\centering
	\footnotesize
	\begin{tabular}{rr}
	\toprule
		& FILO2 vs FILO\\
	\midrule
	$H_0$ & 0.064453 \\
	$H_1$ & 0.032227 \\
		\midrule
	& Similar  \\
	\bottomrule
	\end{tabular}
	\qquad
	\begin{tabular}{rr}
	\toprule
		& FILO2 (long) vs FILO (long)\\
	\midrule
	$H_0$ & \textbf{0.001953} \\
	$H_1$ & \textbf{0.000977} \\
		\midrule
	&  Better\\
	\bottomrule
	\end{tabular}
\end{table}

\clearpage
\subsection{Computational Details for \texorpdfstring{$\mathbb{I}$}{I} Instances} \label{appendix:i_instances}
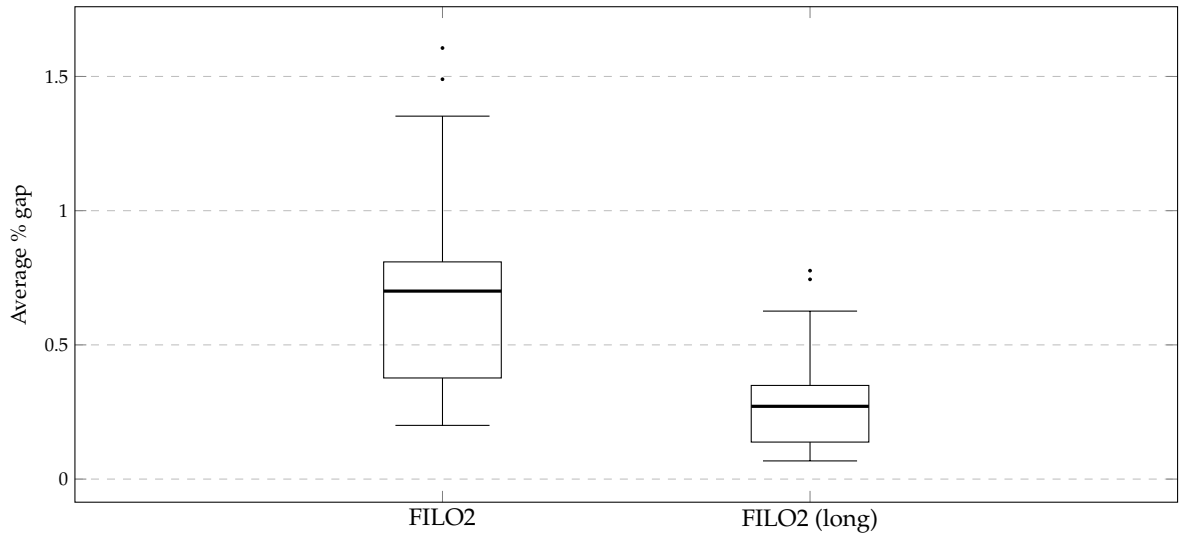
\begin{figure}
	\centering
	\scriptsize
\makeatletter
\pgfplotsset{
    boxplot/draw/average/.code={ 
        \draw [/pgfplots/boxplot/every average/.try]
        \pgfextra
        \pgftransformshift{%
            \pgfplotsboxplotpointabbox
            {\pgfplotsboxplotvalue{average}}
            {0.5}%
        }%
        \endpgfextra
        ;
    }
}
\makeatother
\begin{tikzpicture}
\begin{axis}
[
clip=false,
boxplot/draw direction=y,
boxplot/variable width,
boxplot/every median/.style={black,very thick,solid},
xmin=0.5,xmax=2,
width=1.00\textwidth,
height=0.33\textheight,
ylabel style={align=center}, 
y tick label style={align=right},
x tick label style={align=center},
xtick={1, 1.5},
ymajorgrids=true,
grid style=dashed,
xticklabels={{\small FILO2}, {\small FILO2 (long)}}, xticklabel style={rotate=0}, ylabel={\footnotesize Average \% gap},
scatter/classes={ a={mark=star}, b={mark=*}}]

\addplot[
mark=*, mark size=0.5pt,boxplot,
boxplot prepared={draw position=1, median=0.700369532115902,upper quartile=0.8091757811960527,lower quartile=0.3767139456283748,upper whisker=1.352168825854362,lower whisker=0.20009327505970342,sample size=1}
] coordinates{(1, 1.60598194) (1, 1.48969908)};

\addplot[
mark=*, mark size=0.5pt,boxplot,
boxplot prepared={draw position=1.5, median=0.2710911851285306,upper quartile=0.34891105137254635,lower quartile=0.13788248594617983,upper whisker=0.6258360889562659,lower whisker=0.06770998745464976,sample size=1}
] coordinates{(1.5, 0.77649173) (1.5, 0.74412914)};


\end{axis}
\end{tikzpicture}
	\caption{Comparison of average gaps obtained by algorithms on the $\mathbb{I}$ dataset.}
	\label{figure:i_box_plots}
\end{figure}

Detailed results about computations on the $\mathbb{I}$ dataset can be found in Table \ref{table:i-details} and Figure \ref{figure:i_box_plots}.
\begin{table}
\footnotesize
\centering
\caption{Computations on $\mathbb{I}$ instances.}
\label{table:i-details}
\begin{tabular}{l rrrrrrrrrrr }
\toprule
 & &&  \multicolumn{4}{c}{FILO2} && \multicolumn{4}{c}{FILO2 (long)} \\
\cmidrule{4-7}
\cmidrule{9-12}
ID & BKS && Best & Avg & Worst & $t$ && Best & Avg & Worst & $t$ \\
\midrule
Valle-D-Aosta & 21679514 &&       0.22 &       0.28 &       0.36 &        192 &&       0.08 &       0.13 &       0.19 &       2620\\
Molise & 111184982 &&       0.36 &       0.41 &       0.47 &        146 &&       0.13 &       0.16 &       0.18 &       1901\\
Trentino-Alto-Adige & 102063181 &&       0.81 &       0.92 &       1.08 &        194 &&       0.31 &       0.40 &       0.49 &       2292\\
Basilicata & 175623919 &&       0.59 &       0.74 &       0.92 &        185 &&       0.16 &       0.31 &       0.45 &       2722\\
Umbria & 545507981 &&       0.26 &       0.29 &       0.31 &        248 &&       0.11 &       0.12 &       0.14 &       2124\\
Abruzzo & 311712556 &&       0.66 &       0.73 &       0.80 &        278 &&       0.24 &       0.28 &       0.34 &       2613\\
Friuli-Venezia-Giulia & 415805616 &&       0.54 &       0.60 &       0.70 &        393 &&       0.19 &       0.25 &       0.32 &       4116\\
Liguria & 1426389867 &&       0.18 &       0.20 &       0.22 &        332 &&       0.06 &       0.07 &       0.07 &       2964\\
Calabria & 1964651530 &&       0.36 &       0.40 &       0.42 &        292 &&       0.11 &       0.13 &       0.17 &       2299\\
Marche & 420484426 &&       0.65 &       0.69 &       0.74 &        408 &&       0.22 &       0.26 &       0.31 &       3634\\
Sardegna & 827934149 &&       1.43 &       1.61 &       1.78 &        316 &&       0.61 &       0.78 &       0.97 &       2690\\
Campania & 391859276 &&       0.93 &       1.01 &       1.10 &        420 &&       0.33 &       0.37 &       0.41 &       4417\\
Piemonte & 2627446164 &&       0.30 &       0.32 &       0.35 &        404 &&       0.14 &       0.15 &       0.17 &       2721\\
Toscana & 1084417188 &&       0.69 &       0.77 &       0.85 &        488 &&       0.28 &       0.33 &       0.40 &       5793\\
Puglia & 1464797603 &&       1.36 &       1.49 &       1.59 &        436 &&       0.64 &       0.74 &       0.89 &       2957\\
Sicilia & 1774262462 &&       1.23 &       1.35 &       1.42 &        469 &&       0.50 &       0.63 &       0.75 &       3298\\
Veneto & 1050488613 &&       0.62 &       0.71 &       0.80 &        600 &&       0.25 &       0.29 &       0.34 &       5131\\
Emilia-Romagna & 5405446715 &&       0.22 &       0.24 &       0.25 &        501 &&       0.09 &       0.09 &       0.10 &       3263\\
Lombardia & 1339900081 &&       0.70 &       0.75 &       0.79 &        562 &&       0.29 &       0.34 &       0.39 &       7990\\
Lazio & 3145381332 &&       0.37 &       0.40 &       0.43 &        532 &&       0.12 &       0.14 &       0.15 &       3938\\
\midrule
Mean &  &&       0.62 &       0.70 &       0.77 &        370 &&       0.24 &       0.30 &       0.36 &       3474\\
\bottomrule
\end{tabular}
\end{table}

\clearpage
\printbibliography

\end{document}